\documentclass[11pt]{article}

\usepackage{amsfonts}

\usepackage{amssymb}
\usepackage{latexsym}
\usepackage{graphicx}
\usepackage[english]{babel}

\usepackage{amsfonts}
\usepackage{latexsym}
\usepackage{graphicx}
\usepackage[english]{babel}
\begin{document}
\input epsf

\def\p{\partial}
\def\h{{1\over 2}}
\def\be{\begin{equation}}
\def\bea{\begin{eqnarray}}
\def\ee{\end{equation}}
\def\eea{\end{eqnarray}}
\def\d{\partial}
\def\la{\lambda}
\def\eps{\epsilon}
\def\bb{\bigskip}
\def\mm{\medskip}
\newcommand{\newsection}[1]{\section{#1} \setcounter{equation}{0}}
\def\bb{$\bullet$}
\def\sq{{1\over \sqrt{2}}}
\def\sqi{{1\over \sqrt{2}}}
\def\q{\quad}
\def\z{|0\rangle}
\def\o{|1\rangle}
\def\d{\downarrow}
\def\u{\uparrow}

\def\h{{1\over 2}}
\def\t{\tilde}
\def\r{\rightarrow}
\def\nn{\nonumber\\}

\def\b{\bigskip}

\def\c{C^{(2)}}

\begin{flushright}
%OHSTPY-HEP-T-03-012\\
\end{flushright}
\vspace{20mm}
\begin{center}
{\LARGE  What the information paradox is {\it not}}
\\
\vspace{18mm}
{\bf  Samir D. Mathur }\\

\vspace{8mm}
Department of Physics,\\ The Ohio State University,\\ Columbus,
OH 43210, USA\\mathur@mps.ohio-state.edu
\vspace{4mm}
\end{center}
\vspace{10mm}
\thispagestyle{empty}
\begin{abstract}

There still exist many confusions about the  black hole information paradox and its resolution.  We first give a precise formulation of the paradox, in four steps A-D. Then we examine several proposals for resolving the paradox. We note that  in each case one  of these four steps  has been ignored, so that the proposal does not really target the essence of the paradox. Finally, we give a brief summary of the fuzzball construction and argue that it resolves the paradox in string theory. This resolution contains a deep lesson -- the phase space of quantum gravity is so large that the measure in the path integral can compete with the classical action for macroscopic objects undergoing gravitational collapse.

\end{abstract}
\vskip 1.0 true in

\newpage
\setcounter{page}{1}

\section{Introduction}

A remarkable number of confusions continue to surround the black hole information paradox \cite{hawking}. A section of the general relativity community believes that information will be lost in the process of black hole evaporation. Some say that remnants are left at the end of the evaporation process. Others conjecture that information will emerge in Hawking radiation, but do not offer a concrete mechanism for information recovery; thus the believers in information loss/remnants cannot see  where Hawking's original computation goes wrong.
String theory has shown us the way out of the paradox by an explicit construction of black hole microstates -- fuzzballs -- but relatively few people have studied the nature of this construction, so its implications are not clear to a large section of the black hole community.

The underlying reason behind much of this confusion is that the original paradox found by Hawking is poorly understood today. Almost all physicists have heard of the paradox, but in most cases this knowledge comes through fragments of conversation or tangential remarks in papers. In the process the original arguments laid down by Hawking have become somewhat obscured.  Not surprisingly, it has become hard for physicists to put together the progress in our understanding of black holes and come to a common conclusion about the issue of information loss.  

In this article we will first lay out the Hawking paradox as an argument in four steps A,B,C,D where the last step D is a recently proved inequality that makes Hawking's argument into a `theorem'. This theorem establishes that we will necessarily have information loss/remnants given certain basic assumptions about how gravity and quantum theory work; thus to get information recovery in Hawking radiation {\it we must find a physical process in our theory of gravity that violates one of these assumptions}. We next turn to some ways in which people have attempted to  resolve the paradox, noting that in  each of these cases one of the steps A-D has been {\it ignored}, so that the true problem is not really being addressed. Finally we summarize the fuzzball construction that resolves the paradox in string theory, noting that some common objections to this resolution mistake either the nature of the paradox of the nature or the construction.

\section{The paradox}\label{hawking}

The argument proceeds in the following steps (for more details see \cite{cern}):

\b

\subsection{(A): Existence of a `lab physics' limit} 

The exact description of any physical process must include the effects of quantum gravity. But our experience suggests that there is a separation of scales, so that under suitable conditions we get `lab physics'; i.e., physics described to good accuracy by quantum fields on gently curved spacetime. The paradox starts with listing the commonly accepted conditions under which we must get  `lab physics'; these conditions are termed the `niceness conditions' on spacetime and its slicing:

\b

(N1) Our quantum state is defined on a spacelike slice. The intrinsic curvature ${}^{(3)} R$ of this slice should be much smaller than planck scale everywhere: ${}^{(3)} R\ll {1\over l_p^2}$. 

\b

(N2) The spacelike slice sits in an 4-dimensional spacetime. Let us require that the slice be nicely embedded in the full spacetime; i.e., the {\it extrinsic} curvature of the slice $K$ is small everywhere: $K\ll {1\over l_p^2}$. 

\b

(N3) The 4-curvature  of the full spacetime in the neighbourhood of the slice should be small everywhere ${}^{(4)} R\ll {1\over l_p^2}$

\b

(N4) We should require that all matter on the slice be `good'. Thus any quanta  on the slice should have wavelength much longer than planck length ($\lambda\gg l_p$), and the  energy density $U$ and momentum density $P$ should be small everywhere compared to planck density: $U\ll l_p^{-4}, ~~~P\ll l_p^{-4}$. 
Let us add here that we will let all matter satisfy the usual energy conditions (say, the dominant energy condition). 

\b

(N5) We will evolve the state on the initial slice to a later slice; all slices encountered in the evolution should be `good' as above. Further, the lapse and shift vectors needed to specify the evolution should change smoothly with position: $
{dN^i\over ds}\ll {1\over l_p}, ~~~~{dN\over ds}\ll {1\over l_p} $.

\b

One may wish to add further conditions, but  there are some constraints. First, the conditions should be either commonly accepted as reasonable or be derivable in a complete theory of quantum gravity like string theory. Second, if we impose a new condition, then we will have to  accept its consequences  in {\it all} situations, not just the black hole. 

\subsection{(B): Constructing a `good slicing' of the black hole spacetime}\label{slices}

\begin{figure}[htbp]
\begin{center}
\includegraphics[scale=.55]{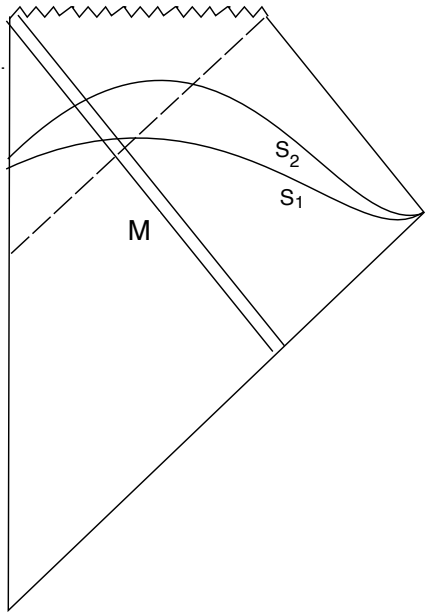}
\caption{{The Penrose diagram of a black hole formed by collapse of the `infalling matter'. The spacelike slices satisfy all the niceness conditions N.}}
\label{fthree}
\end{center}
\end{figure}

 One takes the Schwarzschild black hole metric
\be
ds^2=-(1-{2M\over r}) dt^2+(1-{2M\over r})^{-1} dr^2 + r^2 d\Omega_2^2
\label{one}
\ee
which can be continued past the coordinate singularity at the horizon $r=2M$ to give the Penrose diagram in fig.\ref{fthree}. We will call this the traditional black hole metric. The horizon is a region of smooth gently curved spacetime where the matter fields are in a vacuum state; we call such a horizon a `traditional horizon'. 

\begin{figure}[htbp]
\begin{center}
\includegraphics[scale=.18]{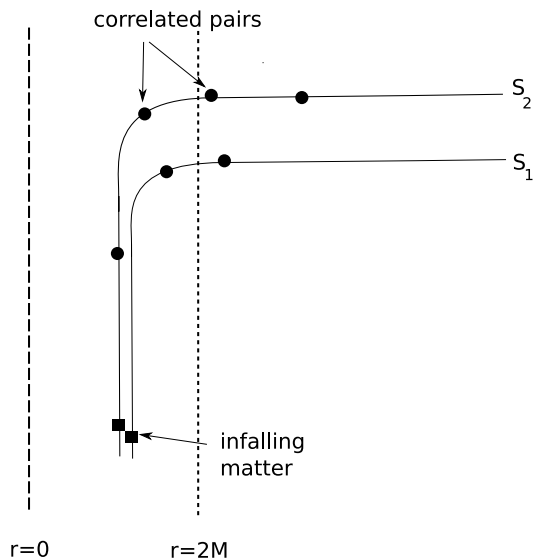}
\caption{{A schematic set of coordinates for the Schwarzschild hole. Spacelike slices are $t=const$ outside the horizon and $r=const$ inside. Assuming a solar mass hole,  the infalling matter is $\sim 10^{77}$ light years from the place where pairs are created, when we measure distances along the slice. Curvature length scale is $\sim 3 ~km$ all over the region of evolution covered by the slices $S_i$.}}
\label{ftwo}
\end{center}
\end{figure}

We now look at a set of `good slices' of this traditional black hole geometry. Note that the slices must extend both outside and inside the horizon to give complete Cauchy surfaces on which we can define the state of our system. While the spacelike geometry of the slices can be seen from the Penrose diagram, it is helpful to also view them in the schematic figure \ref{ftwo}, where we plot $r$ on the horizontal axis and the vertical axis is a schematic `time'. (This `time' $\tau$ will cease to be timelike at some point along the slice, so this diagram is only schematic; we will not use this time for anything that follows.)

The slices are described as follows:

\b

(a) Outside the hole, for $r>3M$,  we take the slice as $t=t_0=const$.

\b

(b) Inside the hole, time and space interchange roles, since the factor $1-{2M\over r}$ changes sign. We take the spacelike slice to be $r=M$. 

\b

(c) We join the two parts (a),(b) by a smooth connector segment ${\cal C}$ that extends from $r=M$ to $r=3M$, and remains spacelike throughout.  

\b

(d) To study the evolution, we must construct a `later' slice:

\b

(i) Outside the hole, we take $t=t_0+M$, where we have advanced the time by $M$, the characteristic scale of the hole. 

(ii) The connector part ${\cal C}$ is not changed at all. 

(iii) We do not advance the part $r=M$, since advancing this part `forward in time' would correspond to moving it to smaller $r$, and we do not wish to go near the singularity at $r=0$.\footnote{In the figures, for the purposes of clarity,  we show the $r=const$ part of the slices as having evolved slightly forward in time towards smaller $r$; we can always allow such an evolution as long as we make sure that we do not get parametrically close to $r=0$.} We can see from fig.\ref{ftwo} that the segment at $r=M$ would have to be extended for a longer distance to meet ${\cal C}$, thus this part of the slice is `stretched'.

\b

At this point one checks a crucial fact: this set of slices satisfies the `nice slice' conditions N1-N5 listed in (A) above. 

\subsection{(C): Creation of the entangled state}

Suppose the black hole was made by the collapse of a shell of mass $M$. At an `early time' spacelike slice we have only the matter shell $|\psi\rangle_M$ on the slice. The evolution to later slices, however, leads to a `stretching' of the slice, as we have seen above. This stretching creates pairs of quanta on the slice, and the explicit form of this creation process is critical to the paradox.  

\begin{figure}[ht]
%%\sidecaption
% For example, with the graphicx style use
\includegraphics[scale=.25]{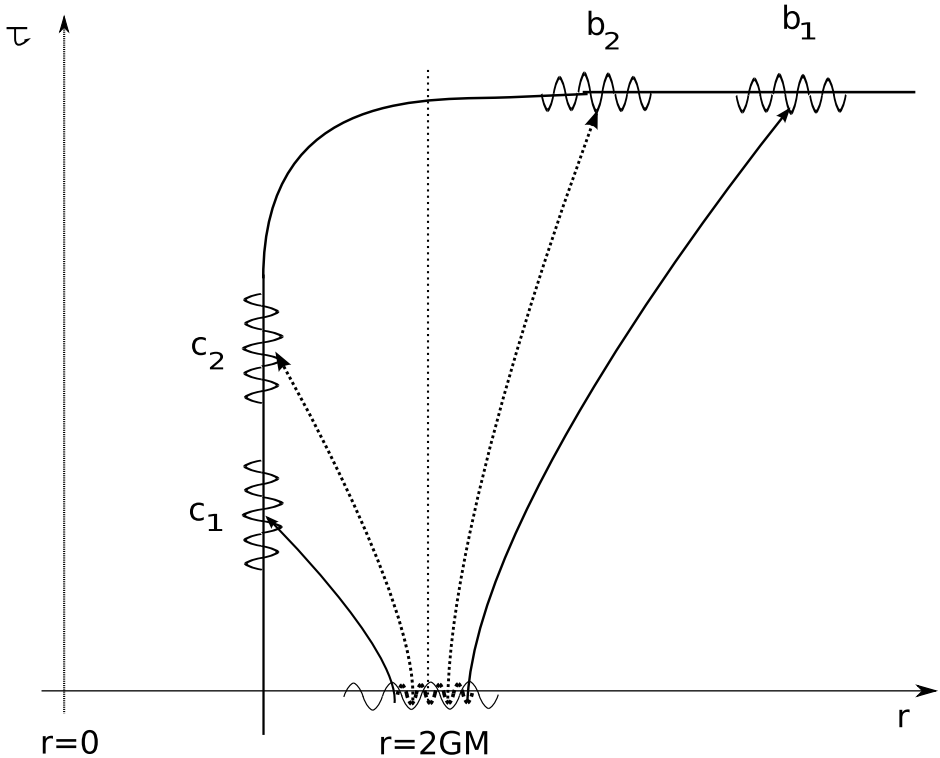}
%
% If no graphics program available, insert a blank space i.e. use
%\picplace{5cm}{2cm} % Give the correct figure height and width in cm
%
\caption{On the initial spacelike slice we have depicted two fourier modes: the longer wavelength mode is drawn with a solid line and the shorter wavelength mode is drawn with a dotted line. The mode with longer wavelength distorts to a nonuniform shape first, and creates an entangled pair $b_1, c_1$. The mode with shorter wavelength evolves for some more time before suffering the same distortion, and then it creates  the entangled pair $b_2, c_2$.}
\label{matftthree}       % Give a unique label
\end{figure}

Define a complete set of particle modes by breaking up the slice into regions of length, say $10M$, and taking wavepackets  of different frequencies in each region. The regions outside the horizon are labelled $b_1, b_2, \dots$, and those inside the horizon are labelled $c_1, c_2, \dots$ (Fig.\ref{matftthree}).   

The initial slice in the foliation has only the matter state $|\psi\rangle_M$ on it, and none of the regions $c_i, b_i$. The first step of evolution stretches the spacelike slice, so that the particle modes $c_1, b_1$ are now present on the new slice. The state of these modes has the schematic form
\be
|\Psi\rangle_{pair}=\sq\Big ( \z_{c_1}\z_{b_1}+\sq\o_{c_1}\o_{b_1}\Big )
\label{three}
\ee
were the numbers $0, 1$ give the occupation number of a particle mode. If we compute the entanglement of the state outside the horizon (given by the mode $b_1$) with the state inside  (given by the mode $c_1$) we obtain 
\be
S_{entanglement}=\ln 2
\ee
At the next step of evolution there is a further stretching, which has two consequences. The modes $b_1, c_1$  at the earlier step move apart, and in the region between them there appears another pair of modes $b_2, c_2$ in a state that has the same form as (\ref{three}). This process repeats, so that after $N$ steps we have the state
\bea
|\Psi\rangle\approx |\psi\rangle_M&\otimes&\Big( \sq \z_{c_1}\z_{b_1}+\sq\o_{c_1}\o_{b_1}\Big)\cr
&\otimes&\Big( \sq \z_{c_2}\z_{b_2}+\sq\o_{c_2}\o_{b_2}\Big)\cr
&\dots&\cr
&\otimes&\Big( \sq \z_{c_N}\z_{b_N}+\sq\o_{c_N}\o_{b_N}\Big)
\label{six}
\eea
The initial matter shell appears in  a simple tensor product with all the other quanta, since the stretching leading to pair creation happens in a region far from where this matter is, and so has no relation to the matter state in the leading order Hawking process. 

The modes $\{ b_i\}$ are entangled with the  $\{M,  c_i\}$ with
\be
S_{entanglement}=N\ln 2
\label{ent}
\ee
This entanglement is seen to grow by $\ln 2$ with each succeeding emission. We now see Hawking's problem: if the hole evaporates away completely, the $b_i$ quanta outside will be in an entangled state, but there will be nothing that they are entangled {\it with}. This means that we can only describe them by a density matrix, so that the initial pure state $|\psi\rangle_M$ has evolved to a mixed state. If the evaporation stops when the hole becomes planck size, then we have a remnant with a large entanglement with the $b_i$. An entanglement entropy $\ln N$ means that the remnant must have at least $N$ internal states, and since we could have started with a hole of arbitrary size, we find that $N$ (and therefore the remnant degeneracy) is unbounded. Having an unbounded number of states for an object with bounded energy and size leads to difficulties with the field theory of such objects. We will lump the possibilities of remnants and information loss together for the purposes of the Hawking argument, since the argument stops when the hole becomes planck scale. 

Even though the simplified form (\ref{six}) holds all the structure needed to understand the information problem, we write the full form of the state for completeness \cite{giddingsnelson}
\be
|\Psi\rangle=\sum_{\{ n\}} C(\{ n\})|\{  n_j\}\rangle_{c_i} 
|\{ n_j\}\rangle_{b_i}, ~~~
C(\{ n\})=C_0\prod_i  \Big (e^{-{1\over 2T}\sum_{j} n_j \omega_j}\Big ) _i
\label{qone}
\ee
Here $|\{  n_j\}\rangle_{b_i} $ denotes the occupation number $n$ state for the $j$th mode in segment $b_i$; this mode has energy $\omega_j$. Similarly,  $|\{  n_j\}\rangle_{c_i} $ denotes the occupation number $n$ state for the $j$th mode of segment $c_i$.  $T$ is the temperature of the hole, which has been taken as constant here; if we wish to take into account the fact that the temperature changes slowly during the evaporation process, then we simply let $T$ be a slowly varying function of the index $i$. 

\subsection{D: Small corrections}\label{d}

This is where Hawking's original analysis stood, but there is an important step that we must add. Since the evolution from one timestep to another has been in a region of gently curved spacetime, we had no choice but to accept that we create a pair like (\ref{three}) at such a step. But there can always be {\it some} corrections to the state (\ref{three}), due to quantum gravity effects that we have not taken into account. All that we are forced to require from validity of the niceness conditions N1-N5 is that these corrections be small.  

Many people believed that these small corrections could have a cumulative effect on the Hawking process that would change his conclusion. After all, the number of emitted quanta is very large ($N\sim \Big ({M\over m_{pl}}\Big )^2$ for a 4-d hole), and the largeness of $N$ might offset the smallness of the corrections in such a way that the net entanglement between the $\{b_i\}$ and $\{M, c_i\}$ becomes zero by the time the hole disappears. Such a reduction of entanglement happens when a piece of paper burns away; at the end we have a pure state of the radiated products $\{b_i\}$ with no entanglement with the location where the paper was initially placed. But on the other hand if the solution of the information puzzle were that simple, then one has to wonder why no one made a simple toy model illustrating how small corrections to the full state (\ref{six}) would lead to a state with no entanglement between the $\{b_i\}$ and $\{M,c_i\}$. 

The situation was clarified by an inequality derived in \cite{cern}, where it was shown that small corrections do {\it not} change the conclusion that the entanglement rises with each emitted pair. The computation  is set up 
as follows. We can choose a basis of orthonormal states $\psi_r$ for the $\{M,c_i\}$ quanta inside the hole, and an orthonormal basis $\chi_s$ for the quanta $\{b_i \}$ outside the hole, such that the state at timestep $n$ can be written as
\be
|\Psi_{M, c, b}(t_{n})\rangle=\sum_{r,s} C_{rs} \psi_r\chi_s
\ee
It is  convenient to make unitary transformations on the $\psi_r, \chi_s$  so that we get
\be
|\Psi_{M, c, b}(t_{n})\rangle=\sum_{i} C_{i} \psi_i\chi_i
\label{stateone}
\ee
In the leading order  evolution we would have  at time step $t_{n}$:
\be
|\Psi_{M, c, b}(t_{n})\rangle\r |\Psi_{M, c, b}(t_{n})\rangle\otimes \Big [\sqi |0\rangle_{c_{n+1}}|0\rangle_{b_{n+1}}+\sqi |1\rangle_{c_{n+1}}|1\rangle_{b_{n+1}}\Big ]
\label{leading}
\ee
where the term in box brackets denotes the state of the newly created pair. 

Let us now write down the modifications  to this evolution from timestep $t_n$ to timestep $t_{n+1}$ that will encode  the small corrections we wish to allow.
 For the quanta that have left the hole at earlier timesteps, we have no change
\be
\chi_i\r \chi_i
\label{chifixed}
\ee
 In the leading order evolution there was no change to the state of $\{ M,c_i\}$ inside the hole, since the part of the spacelike slice carrying these quanta did not evolve forwards in time. We will now allow a completely general unitary evolution of this state to any other state formed by these quanta, since we do not need to constrain what happens in the hole to arrive at the inequality we desire. The essential constraint is on the pair creation process at the next  timestep. 
In the leading order Hawking computation the state of the newly created pair was $|\Psi\rangle_{\rm pair}$ (eq. (\ref{three})). We now allow the state of this pair to lie in a 2-dimensional subspace, spanned by $|\Psi\rangle_{\rm pair}\equiv S^{(1)}$ and an orthogonal vector $S^{(2)}$:
\bea
S^{(1)}&=&\sqi  |0\rangle_{c_{n+1}}|0\rangle_{b_{n+1}}+\sqi  |1\rangle_{c_{n+1}}|1\rangle_{b_{n+1}}\cr
S^{(2)}&=&\sqi  |0\rangle_{c_{n+1}}|0\rangle_{b_{n+1}}-\sqi  |1\rangle_{c_{n+1}}|1\rangle_{b_{n+1}}\cr
&&
\label{set}
\eea
Thus the most general evolution allowed for the state in the hole is 
\be
\psi_i\r \psi_i^{(1)}S^{(1)}+\psi_i^{(2)}S^{(2)}
\ee
where $\psi_i^{(1)}, \psi_i^{(2)}$ are any two states of $\{M,c_i\}$ (the initial matter and the infalling members of Hawking pairs produced at earlier steps).  Unitarity of evolution gives
$
|| \psi_i^{(1)}||^2+|| \psi_i^{(2)}||^2=1
$.
We thus get the evolution
\bea
|\Psi_{M, c, b}(t_{n+1})\rangle&=&\sum_{i} C_{i} [\psi_i^{(1)}S^{(1)}+\psi_i^{(2)}S^{(2)}]~ \chi_i\nn
&=&\Big [ \sum_{i} C_{i} \psi_i^{(1)} \chi_i\Big ]S^{(1)} + \Big [ \sum_{i} C_{i} \psi_i^{(2)} \chi_i\Big ]S^{(2)}\nn
& \equiv &  \Lambda^{(1)}S^{(1)}+  \Lambda^{(2)}S^{(2)}
\label{qthree}
\eea
Since $S^{(1)}, S^{(2)}$ are orthonormal, normalization of $|\Psi_{M, c, b}(t_{n+1})\rangle$ implies that
$
||\Lambda^{(1)}||^2+||\Lambda^{(2)}||^2=1
$.
We can now state precisely what it means for the quantum corrections to be small. If the evolution from step $n$ to step $n+1$ is to be close to the semiclassical one in any sense then we must have produced mostly the state $|\Psi\rangle_{\rm pair}\equiv S^{(1)}$ and only a small amount of $S^{(2)}$. Thus we write
\be
|| \Lambda^{(2)}||<\epsilon, ~~~\epsilon\ll 1
\label{cond1}
\ee
If there is no such bound, then we will say that the corrections to the Hawking evolution are `order unity'. 

The result of \cite{cern} says that with the smallness condition (\ref{cond1}) we get a minimal increase of entanglement at each step
\be
S_{n+1}-S_n>\ln 2-2\epsilon
\label{result}
\ee

\subsection{Fallacies}

There are three common types errors that one might make in searching for solutions of the information paradox. The first is that we ignore the explicit evolution in step C, and replace it by a new evolution, without giving a mechanism that would achieve this change in our theory of gravity. Thus if we  replace (\ref{six}) (or its more exact version (\ref{qone})) by a different evolution of ones choosing, then we are ignoring Hawking's explicit computation. But the only way we can alter Hawking's entangled state is by showing that the black hole does not have the geometry discussed in step B. If we wish to use such an alternative solution to the black hole then we have to {\it show} that this solution exists in our theory of gravity; we cannot just assume its existence.

The second type of error comes from changing the state on the initial slice in ways that are not permitted by our theory of gravity. We have a smooth slice in a smooth spacetime. All states on such a slice are assumed to be understood. We cannot therefore write an arbitrary set of properties for the state on this slice; for example  we cannot assume that the state is populated by a large number of high frequency quanta without explaining how (in our theory of gravity) these quanta would not have a $T_{\mu\nu}$ that would change the geometry (\ref{one}). 

The third type of error consists of taking a reasonable state on the initial slice, but then assuming a rule for evolution that is not explicitly known to be true in our theory of gravity. One instance that we will discuss is the idea that there is a large `boost' between quanta deep in the hole and quanta near the horizon, and this boost gives rise to large nonlocal effects that completely alter the state (\ref{three}). There is indeed such a boost, but concrete computations have so far failed to reveal any large effect on the pair state (\ref{three}) created   at the horizon.

In short, we must use our gravity theory explicitly to find any deviations from the Hawking argument; small corrections (which are always present) do not change the argument (as noted in step D), and large changes need us to find some explicit gravitational effect that was not  noted in early work with black holes. 

\section{The Rubic cube model}\label{secfour}

As we noted in section \ref{d}, there is no information paradox when we burn a piece of paper; the emitted quanta form a pure state by themselves and there is nothing left behind where the paper was. In \cite{cern} it was explained that that crucial feature characterizing the black hole is  that each stage of evolution creates the {\it same} entangled state (\ref{three}), upto small corrections. The paper, on the other hand, creates a radiated state that depends on the state of atoms on its surface, and this state differs from one emitted quantum to the next. 

To be concrete, let us make a model of burning paper. We start with a set of spins
\be
|\Psi\rangle_0=\d\u\d\u\u\d\d\u\d\u\d
\label{ex4}
\ee
If the rightmost two spins are both down, we emit a down spin, leaving the other down spin behind. if they are both up, we emit an up spin, leaving the other up spin behind.  The emitted spin in each case is written to the right of a vertical bar. We also add rules for the cases where  the rightmost two spins are up-down or down-up:
\bea
\d\u\d\u\u\d\d\u\d\d\d&\r& \d\u\d\u\u\d\d\u\d\d\Big |\d \nn
\d\u\d\u\u\d\d\u\d\u\u&\r& \d\u\d\u\u\d\d\u\d\u\Big |\u \nn
\d\u\d\u\u\d\d\u\d\d\u&\r& \sq\d\u\d\u\u\d\d\u\d\d\Big |\u+ \sq\d\u\d\u\u\d\d\u\d\u\Big |\d\nn
\d\u\d\u\u\d\d\u\d\u\d&\r& \sq\d\u\d\u\u\d\d\u\d\d\Big |\u- \sq\d\u\d\u\u\d\d\u\d\u\Big |\d
\label{qtwo}
\eea
This is a unitary evolution, in which all the spins are emitted one by one, leaving nothing behind. (We can also make a permutation of the spins remaining after each emission; such a permutation was included in the `burning paper' model analyzed in \cite{plummat}.) The entanglement entropy of the emitted spins with the remaining spins first goes up, and then after about the halfway point, comes down, ending at zero. This is in accord with the expected behavior of emission from all normal bodies, as noted by Page \cite{page}.

But the evolution (\ref{qtwo}) cannot be mapped to the evolution (\ref{qthree}) with the smallness condition (\ref{cond1}). The difference is also easy to spot. In the evolution (\ref{qthree}) of the black hole, an entangled pair is created {\it in the same state} (to leading order) at each step; this is because the modes $b_i, c_i$ populated at step $i$ stretch out from unpopulated modes of the vacuum, and the excitation created in  these modes is completely determined by the  rule for propagation of quantum fields on the metric (\ref{one}). The evolution (\ref{qtwo}), on the other hand, gives an entanglement at each step that depends on the state of the two rightmost spins at that timestep.

Thus writing down an evolution (\ref{qtwo}) has no bearing on the information paradox; we have not tried to make any connection to the black hole at all. All we have made is a model of `burning paper', and their are an infinity of such models possible, one for each ordinary body that can be `burnt'. 

\subsection{The Rubik cube model}

Recently a model of black hole evaporation was proposed where we would end up with no entanglement at the end of the evaporation process \cite{rubic}. How could this be true, given that we have a rigorous inequality (\ref{result}) which says that entanglement rises at each step of the emission process? The answer is simple: the model proposed in \cite{rubic} is actually a model of `burning paper', since it does not follow the conditions required for evaporation in the traditional black hole. Let us look at the model of \cite{rubic} in some detail, to identify this issue more clearly.

In this model we imagine the black hole to be made of a large number of Rubik cubes. Each Rubik cube can be in several different states; we will call these states $C_1, \dots C_n$. The state of any cube can be changed by some basic `moves', called $L, R, U$. (These symbols stand for Left, Right, Upper, respectively, but we will not need those details of the model here.) We define  an additional operation $N$ which corresponds to `no change'. 

At each step in the evolution, these operations act on the cubes making up the hole (in a way to be specified shortly), and a particle is emitted in a certain state. Thus we may choose the evolution to correspond to a sum of operators $\vec N, \vec L, \vec R, \vec U$ with equal amplitude, and the emitted particle states will be $n, l, r, u$ respectively. (Here $n$ stands for `no emission', which is one of the states possible at that step of emission; we had termed this possibility $|0\rangle_b$ in  (\ref{three}).) 

The only other rule we have is that if a particular Rubik cube moves to its `solved' state, then it disappears from the list of cubes making up the hole, and a quantum in state $q$ is emitted to account for the energy lost 
from the hole by deletion of the cube. Thus the evolution has the form
\bea
\{ C_1, \dots C_m\}&\r& \h \Big (\{\vec N C_1, \dots \vec N C_m\} |n\rangle +\{\vec L C_1, \dots \vec L C_m\} |l\rangle\nn
&&+\{\vec R C_1, \dots \vec R C_m\} |r\rangle +\{\vec U C_1, \dots \vec U C_{m-1}\} |qu\rangle  \Big )
\eea
where we have assumed that the operation $\vec U$ brought the last cube in state $C_m$ to the `solved' state and hence lead to emission of an extra quantum $q$.

The authors of \cite{rubic} checked carefully that the overall evolution is unitary, and that it led to an eventual disappearance of the hole, with all the energy appearing in the radiated quanta. The final state is a linear combination of states of the form $|unlrqu\dots\rangle$ where we have a general sequence of quanta in the different allowed states. This final state is a pure state by itself, not entangled with anything else, and we seem to have obtained black hole evaporation with no remnants/information loss. Where did the theorem (\ref{result}) go wrong?

The answer, however is clear; this Rubik cube model is like the burning paper model, and not a model of the black hole. Any model of the black hole has to start with a metric that can be demonstrated to exist in the theory (this is usually the metric (\ref{one})), and then has to use a rule for evolution that is forced by the evolution of fields in that metric. In particular, if we assume the metric (\ref{one}) as a good approximation to the state of the hole, then the evolution has to have a structure like (\ref{qthree}) with condition (\ref{cond1}).

Let us write an evolution in the Rubik cube setup that {\it would} accord with these constraints. As the analog of (\ref{three}) we take
\bea
\{ C_1, \dots C_m\}&\r& \h   \{ C_1, \dots C_m\}\otimes {1\over 2}\Big ( n'_1\otimes n_1 + l'_1\otimes l_1 + r'_1\otimes r_1 + u'_1\otimes u_1\Big )
\label{qfour} 
\eea
where the quanta $n', l', r', u'$ are the members of the created pairs inside the hole. The essential feature of the evolution (step C in the Hawking argument) is now captured, and the entanglement between the inside of the hole and the outside has increased (by $\ln 4$ in this case). 
At the next step we will have another increase in entanglement
\bea
\{ C_1, \dots C_m\}&\r&    \{ C_1, \dots C_m\}\otimes {1\over 2}\Big ( n'_1\otimes n_1 + l'_1\otimes l_1 + r'_1\otimes r_1 + u'_1\otimes u_1\Big ) 
\nn
&&~~~~~~~~~~~~~~~~~~~~~~~~\otimes {1\over 2}\Big ( n'_2\otimes n_2 + l'_2\otimes l_2 + r'_2\otimes r_2 + u'_2\otimes u_2\Big ) 
\label{ten}
\eea
Note that we do not have to keep the internal quanta $n'_1, l'_1, r'_1, u'_1$ unchanged at this second step; they were left unchanged in the leading order Hawking process but in our derivation of the inequality (\ref{result}) we have allowed such quanta to be mixed with the state $\{ C_1, \dots C_m\}$ of the hole in an arbitrary way. Thus we can replace (\ref{ten}) by
\bea
\{ C_1, \dots C_m\}&\r&    \{ C'_1, \dots C'_{m'}\}
\otimes {1\over 2}\Big ( n'_2\otimes n_2 + l'_2\otimes l_2 + r'_2\otimes r_2 + u'_2\otimes u_2\Big ) 
\label{tenp}
\eea
where the map
\be
    \{ C_1, \dots C_m\}\otimes {1\over 2}\Big ( n'_1\otimes n_1 + l'_1\otimes l_1 + r'_1\otimes r_1 + u'_1\otimes u_1\Big ) \r  \{ C'_1, \dots C'_{m'}\}
    \ee
is unitary. Thus in particular the index $m'$ would need to be larger than $m$, to accommodate the needed degrees of freedom. What we cannot do is alter the factor ${1\over 2}\Big ( n'_2\otimes n_2 + l'_2\otimes l_2 + r'_2\otimes r_2 + u'_2\otimes u_2\Big ) $ in (\ref{tenp}), since it is forced by evolution of modes in the metric (\ref{one}).   (More precisely, we can allow `small corrections' to this factor in line with (\ref{cond1}), but these corrections do not change the conclusion about monotonically increasing entanglement, and anyway such small corrections were not considered in \cite{rubic}.)

Thus while the the model of \cite{rubic} is a nice model of unitary evolution, it does not build in step C of the Hawking argument, and is therefore not addressing the Hawking paradox at all.

\subsection{Amplitudes vs probabilities}

We still need to ask {\it why} the Rubik cube model was considered as a potential model for black holes.  The reason can be traced to an interpretation of  the term `information free horizon' that does not incorporate its full significance.

The term `information free horizon is meant to signify that the state of quantum fields at the horizon is the natural vacuum arising in the process of gravitational collapse -- this is the state used in Hawking's evolution, and it generates the entangled state (\ref{six}).  It can be seen that the probability of occupation number $0$ in a mode $b_i$ is $\h$, and the probability of occupation number $1$ is $\h$. Thus one can say that for an observer outside the hole, at each step the possibilities of no emission ($0$) or getting an emission ($1$) are randomly selected with equal probability.

{\it But this probability distribution has nothing to do with the meaning of the phrase `information free horizon'.} An `information free horizon' generates the particular entangled state (\ref{six}), not any other state that may have the same probability distribution for observers outside the hole. Thus in particular consider a state like
\bea
|\Psi\rangle\approx |\psi\rangle_M&\otimes&\Big( \sq \z_{c_1}\z_{b_1}-\sq\o_{c_1}\o_{b_1}\Big)\cr
&\otimes&\Big( \sq \z_{c_2}\z_{b_2}+\sq\o_{c_2}\o_{b_2}\Big)\cr
&\dots&\cr
&\otimes&\Big( \sq \z_{c_N}\z_{b_N}-\sq\o_{c_N}\o_{b_N}\Big)
\label{sixp}
\eea
where we have changed some positive signs to negative. This state also has probability $\h$ for emission at each step, but is {\it not} the state created by evolution in an `information free horizon'. 

We can see this issue again in `burning paper' model (\ref{qtwo}) that we constructed above. The probability of emitting an up or a down spin at each step is $\h$. But we should not confuse this with emission from an `information free horizon': as we had noted, the evolution rule (\ref{qtwo})  does not map to a rule which would give state like (\ref{six}).

We can now see that the Rubik cube model of \cite{rubic} has a {\it probability distribution} of emission that could agree with black hole emission, {\it but not the state generated by an information free horizon}. The latter kind of state would be something like (\ref{qfour}), and such an evolution would have Hawking's problem of ever increasing entanglement. 
To understand the paradox we have to focus on the entangled state of the {\it pair} that is created. One member of this pair is outside the horizon and one inside, so we cannot understand the paradox if we focus on reproducing the probabilities seen by the outside observer. 

To summarize, the model of \cite{rubic} is an excellent example of what black holes would do {\it if} their internal structure were such that Hawking radiation was a unitary process. The problem is that the metric (\ref{one}) does not give an evolution like the one used in \cite{rubic}, and when people tried to find deformations of  this metric, they did not succeed -- this is the `no-hair theorem'. As we will see later, the fuzzball construction {\it did} find the needed `hair', and then black holes do emit like burning paper.

\section{Hawking radiation as tunneling}

There are many derivations of the Hawking radiation spectrum, apart from Hawking's original one. One of these is the method of `tunneling', where one considers the path of a  shell that starts as a virtual excitation but then tunnels its way out of the horizon to escape as a real quantum. This method allows computation of `backreaction' in the sense that the hole left behind has energy $M-\Delta E$, where $\Delta E$ is the mass of the escaped shell \cite{kraus,paddy,parikh}. 

It has been noted  that the Hawking emission rate arising in this method of computation can be written as
\be
\Gamma(\Delta E)\sim e^{-\Delta S_{bek}}
\label{pw1}
\ee
where $\Delta S_{bek}$ is the decrease in the Bekenstein entropy \cite{bek} of the hole upon reduction of its mass by $\Delta E$. It has been argued  in \cite{pw2} that the changes in temperature upon each emission (computed from the backreaction) generate correlations among the emitted quanta and remove the information problem. 

There are several things wrong with this argument, but before discussing the information question we begin with a few basic observations. The relation (\ref{pw1}) has a $\sim$ sign instead of an equality. This is standard in leading order tunneling calculations, where the exponent arises from the classical action, and the prefactor is an order $\hbar$ correction. But in the present case $\Delta S_{bek}\sim 1$ for emission of a typical quantum with energy $\Delta E$ of order the temperature $ T$:
\be
\Delta S_{bek}\sim M\delta M\sim {1\over T}{\Delta E}\sim 1
\ee
Thus the exponent on the RHS in (\ref{pw1}) is order unity. But in that case we have to ask:  what is the significance of the estimate (\ref{pw1}) if we dont know the prefactor? The part we are missing is of the same order as the part we have written. Further, there is some essential physics in the missing part, since  (\ref{pw1}) has different units on its two sides: $\Gamma$ is the probability of emission {\it per unit time}, while the RHS has no units.

This missing prefactor may appear to be a minor irritant, since one could try to perform a higher order computation and get the prefactor. But things are not so simple. In a tunneling computation like $\alpha$-decay, the exponent comes from the WKB action under the barrier, and the prefactor is computed from the number of times the tunneling $\alpha$-particle hits the barrier per second as it bounces back and forth in its potential well. Thus this prefactor has units of ${1\over {\rm time}}$, and supplies the correct units for $\Gamma$. But how do we think of the black hole interior as a box in which the shell was bouncing back and forth before tunneling out? 

Addressing this question will bring us to the essence of the information problem. The $\alpha$-decay example yields a lesson for  the black hole case: to understand the prefactor we will need to know the wavefunction of the shell in the {\it interior} of the horizon. This  is exactly the part of the wavefunction which in the Hawking computation gets entangled with the part of the wavefunction outside. If we ignore this interior part of the wavefunction, we miss the entanglement completely, and thus fail to understand the paradox. 

Thus we see that there is no way to proceed with an analysis of the information problem on the basis of a relation like (\ref{pw1}): this estimate of the radiation rate has no data about the nature of the entanglement. We now wish to consider a computation presented in  \cite{pw2} where it was argued that emissions with the probability structure (\ref{pw1}) result in a complete recovery of the information of the hole. Before we note the argument, it will be helpful to consider  a simpler process than  black hole emission: Schwinger pair production in an electric field. Thus consider two parallel plates, the left one carrying a charge $Q$ and the right one a charge $-Q$. We can think of the left plate as analogous to the interior of the hole and the right plate as the exterior of the hole. The field between the plates creates $e^+e^-$ pairs, with the $e^-$ flying towards the left plate and the $e^+$ flying towards the right plate. Let us collect these $e^+, e^-$ quanta in some detectors placed just before the plates, so that we can measure their state later at leisure. Let the spins of the  $e^+e^-$ be entangled, so that the overall state is a singlet
\be
\sq \Big ( |\uparrow\rangle_{e^-}|\downarrow\rangle_{e^+}-|\downarrow\rangle_{e^-}|\uparrow\rangle_{e^+}\Big )
\ee
This entanglement is similar to  the Hawking radiation case, and the left and right sides of our experiment are entangled with entangled entropy $\ln 2$. 
Succeeding emissions will increase this entanglement,  by $\ln 2$ at each timestep. 

Now suppose somebody argues in the following way. ``At the first timestep the state of the $e^+$ on the right had an entanglement entropy $\ln 2$. Thus collecting the $e^+$ brought an entropy $\Delta S=\ln 2$ to the right side. This means that there has been a decrease in entropy by $\ln 2$ on the left side. After $N$ steps, we have removed an entropy $N\ln 2$ from the left side, so the entropy on the left should have gone down by $N\ln 2$.''

This argument is of course manifestly incorrect; entropy arises from the entanglement of the two sides, and is not a quantity that moves out of one side to the other. The entropy of the left side has not decreased; it has in fact increased with each emission. 

Now let us summarize the argument of \cite{pw2} for the black hole case:

\b

(a) We write $P[E]=\Gamma[E]$, identifying the probability $P[E]$ of emitting a quantum of energy $E$ with the rate of Hawking emission $\Gamma[E]$. This identification has the problem noted above that that the two quantities have different units. 

(b) We note that the shell tunneling calculation gives
\be
\Gamma[E]\sim e^{-8\pi E(M-{E\over 2})}
\label{gamma}
\ee
We ignore the difference in units between various quantities, and write
\be
P[E]=\Gamma[E]=e^{-8\pi E(M-{E\over 2})}=e^{-\Delta S_{bek}}
\label{inew}
\ee
where in the last step we have used $S_{bek}=4\pi M^2$. 

(c) We make two observations about the function $\Gamma[E]$ in (\ref{inew}) :

(i) We have
\be
\Gamma[E_1,E_2]=\Gamma[E_1]\Gamma[E_2 |E_1]=\Gamma[E_1+E_2]
\label{ineqq}
\ee
where  $\Gamma[E_2|E_1]$ is the probability of emitting $E_2$ given that $E_1$ has already been emitted (and thus lowered $M$ to $M-E_1$). Note, however, that this relation is true even if we ignore the backreaction due to emission of the first quantum, and just take 
\be
\Gamma=e^{-8\pi EM}
\label{iner}
\ee

(ii) We note that
\be
\Gamma[E_1, E_2]\ne \Gamma[E_1]\Gamma[E_2]
\label{ineq}
\ee
if we use $\Gamma[E]=e^{-8\pi E(M-{E\over 2})}$. Here we would have obtained $\Gamma[E_1, E_2]= \Gamma[E_1]\Gamma[E_2]$ if we had used $\Gamma=e^{-8\pi EM}$. 

There is an important issue to note here. It has been argued in some of the articles in \cite{pw2} that the inequality (\ref{ineq}) is responsible for the correlations that transfer information to the radiation. But as we will see, the actual computation uses only the relation (\ref{ineqq}), which holds even for $\Gamma=e^{-8\pi EM}$ where we have {\it no} such correlations. 

(d) The probability for emitting a sequence of quanta $E_1, \dots E_n$ leading to complete evaporation of the hole is 
\bea
P[E_1, \dots E_n]&=&\Gamma[E_1, \dots E_n]
=\Gamma[E_1]\Gamma[E_2|E_1]\dots \Gamma[E_n|E_1,\dots E_{n-1}]\nn
&=&\Gamma[E_1+\dots +E_n]=\Gamma[M]=e^{-S_{bek}}
\eea 
where at the last step we have used (\ref{ineqq}). Letting each emitted sequence correspond to a different initial microstate of the hole, we find  that there were $e^{S_{bek}}$ possible initial microstates. Since this equals the Bekenstein entropy of the hole, we conclude that the  emitted particles have carried away the entropy of the hole, and there is no information loss.

\b

It is easy to see where this argument for information loss fails. As we already noted, the mismatch of units in (\ref{inew}) hides the important fact that we are missing the part of the wavefunction  that is inside the hole.\footnote{The computation of \cite{kraus} starts with the full wavefunction on the Cauchy slice, and then proceeds to get the emission rate. In such a computation the entanglement of the wavefunction on the two sides of the horizon can in principle be computed; these papers do not argue that correlations inherent in (\ref{pw1}) will lead to a resolution of the information paradox. The paper \cite{paddy} does not argue for information recovery either. Thus the difficulty is not the the tunneling computation itself; the difficulty lies with the suggestion that correlations inherent in the tunneling calculation will resolve the information problem. } The argument that (\ref{ineq}) builds in information-carrying correlations seems to have no relevance to the final result, since (\ref{ineqq}), the equation we actually use, is true even with the choice (\ref{iner}) which has no such correlations. Most importantly, step (d) makes the same fallacious argument that we noted for the Schwinger paradox: it {\it assumes} that the black hole radiates like burning paper and each emitted quantum carries entropy {\it away} from the  hole. As we have seen above, such is not the case; the Hawking problem arises because of a growing entanglement entropy between the inside and outside of the hole. The paradox can be addressed only when we start with the full entangled wavefunction which straddles both sides of the horizon; the paradox cannot be addressed by focusing only on the probabilities of emission.

A final comment: when people describe the Hawking process in terms of probabilities, they sometimes say that information will be contained in `deviations from thermality'. We must be careful with such statements, however. The `thermal spectrum' is a particular {\it probability distribution}, and as we have remarked above, not directly related to the entangled wavefunction. As a simple illustration of the difference, note that the wavefunction (\ref{qone}) has a thermal distribution of emitted quanta, while the simplified wavefunction (\ref{six}) has all its emission at just one frequency (which we can imagine to be $\omega=T$). The amount of entanglement at each step is however of the {\it same} order in the two cases. Thus the essence of the Hawking problem -- the entanglement -- has very little to do with the distribution of emitted frequencies. 
%%%%%%%%%%%%%%%%%%%%%%%%%%%%%%%%%%%%%%%%%%%%%%%%%%%%%%%%%%%%

\section{Difficulties with modifying the entangled state (\ref{qone})}

In section \ref{secfour} we saw that writing an arbitrary unitary evolution for the black hole is incorrect: to address the paradox we have to {\it find a reason} why the specific evolution (\ref{qthree}),(\ref{cond1}) should be modified. We now move on to attempts to modify this evolution by finding suitable effects in quantum gravity. We can look for modifications in the state of the system on the initial slice, or we can look for modifications in the evolution rule created by  specific quantum gravity interactions. 

\subsection{The creation of pairs}

Let us recall where the entangled state (\ref{qone}) comes from. Suppose we have a massless scalar field. We quantize  by expanding in modes
\begin{equation}
\hat\phi(x)=\sum_n \left (\hat a_n f_n(x)+\hat a^\dagger_n f_n^*(x)\right  )
\label{fn}
\end{equation}
where the mode functions $f_n(x)$ satisfy the wave equation $\square f_n=0$. Imposing $[\hat a_n, \hat a^\dagger_m]=\delta_{n,m}$, we define a vacuum state  $\hat a_n|0\rangle_a=0$, and particle excitations are then given by acting with $\hat a^\dagger_n$ on $|0\rangle_a$. 

When we consider a later slice in the evolution, different field modes may be natural for defining particles. Writing
\begin{equation}
\hat\phi(x)=\sum_n \left (\hat b_n g_n(x)+\hat b^\dagger_n g_n^*(x)\right  )
\label{fnq}
\end{equation}
with $\square g_n=0$, $[\hat b_n, \hat b^\dagger_m]=\delta_{n,m}$, we define a new vacuum by $\hat b_n|0\rangle_b=0$, and get particles created by $\hat b^\dagger_n$ on $|0\rangle_b$. 

The Hawking computation consists of taking the slicing in step B of section \ref{hawking} and finding that a vacuum state of fields on an initial slice evolves to a state of the form (\ref{qone}) after evolution. It may appear at first glance that particle creation is not a very well-defined process, since different choices of modes lead to different numbers of particles on a slice. But it should be noted that there {\it is}  a standard definition of particles for the modes $b_i$ which are in the part of the slice outside the horizon. One may use different choices of modes inside the horizon;  this region of spacetime is curved with curvature of the same order as the wavelength of the created quanta, and no particular choice is canonical.  Taking one of these choices,  the form (\ref{qone}) was obtained in \cite{giddingsnelson}. But the important point is that {\it whatever basis we use for the modes inside the horizon, the entanglement entropy between the $b_i$ modes outside and the modes inside will be the same.} This follows from a simple mathematical fact: if two systems are entangled, and we make a unitary transformation on the states in one of the systems, then the entanglement entropy between the two systems is not altered.

\subsection{Adding extra quanta on the initial slice}

To resolve the information problem, people tried to change the entangled state (\ref{six}) to a different state, which would have {\it no} entanglement between the parts inside and outside the hole. (We will use the schematic form (\ref{six}) in place of the full state (\ref{qone}) for simplicity in the following discussion.) The first place to look for changes would be to consider alternative states on the initial slice of the foliation.

We have seen that if the first slice has the vacuum state in its `middle' part (the neighborhood of the part that will stretch), then the next slice in the evolution has the particular entangled state  $\sq\Big ( \z_{c_1}\z_{b_1}+\sq\o_{c_1}\o_{b_1}\Big )
$. Thus suppose the middle part of the first slice was not in the vacuum state but had a particle on it, with  wavelength $\lambda\sim R$, where $R$ is the Schwarzschild radius of the hole. All lengths scales in the evolution are determined by $R$: the size of the region that stretches, the time gap between the two slices, and the wavelengths of the created $b_i, c_i$ quanta are all $\sim R$. 
Thus the extra quantum we have placed on the initial slice can affect the state on the later slice by order unity; for example if we are looking at  a bosonic field, then we get a bose enhancement for creating a quantum if a quantum was already present in that mode. Let us assume that the state of the created pair changes as
\be
\sq\Big ( \z_{c_1}\z_{b_1}+\sq\o_{c_1}\o_{b_1}\Big )~\r~{1\over \sqrt{5}}\Big ( \z_{c_1}\z_{b_1}+2\o_{c_1}\o_{b_1}\Big )
\label{qfive}
\ee
We have now succeeded in modifying the Hawking evolution by order unity at this step. This state is still entangled between the $b_1$ and $c_1$ modes, but at least we have moved away from the Hawking state (\ref{three}) by order unity, and can hope a succession of such changes might yield a state with zero entanglement. 

 But unfortunately  such changes do not persist at the {\it next} evolution step. The evolution of slices is a  process of progressive `stretching'. The extra quantum we had placed on the initial slice, and the quanta created at step (\ref{qfive}), move away from the `creation region' and the next quantum is again created from the vacuum in the state
\be
\sq\Big ( \z_{c_2}\z_{b_2}+\sq\o_{c_2}\o_{b_2}\Big )
\ee
and so on for all future steps. We have not solved the Hawking problem this way.

Why dont we put {\it many} quanta on the initial slice, so that the pair state created at each step is modified by order unity? Since there is a stretching at each timestep, one finds that on the initial slice one needs to add quanta with wavelengths $R, R/2, R/4, \dots$.  With such a state, we can imagine that at each timestep there is a quantum present that will affect the process of pair creation by order unity.  But how many such quanta will we need to take?

The hole emits $\sim \Big ( {M\over m_{pl}}\Big )^2$ quanta in its evaporation process. If we try to populate all the relevant modes on the initial slice, then we will have to populate modes of energy much larger than planck energy, and the geometry in the vicinity of the slice will be nowhere near the metric (\ref{one}). It is unclear how such a state can be created in the gravitational collapse of a smooth, finite density shell.

A final possibility is that we can try to populate only a few modes, say the first 100 in the set with wavelengths $R, R/2, R/4 \dots$, and trust that when these modes stretch away, new excitations will be created on the slice to take their place. In that case we would have a `gas' of quanta floating in the vicinity of the horizon, modifying the evolution away from the Hawking state (\ref{six}). 

But the `no-hair theorem' tells us that we cannot find such states \cite{nohair}. The `theorem' is not really a rigorous result, but a statement of the fact that considerable work has failed to find a deformation of the metric (\ref{one}) that maintains its structure over time. That is, we can certainly add excitations to the black hole metric on one time slice, but as we follow the evolution in the good slicing, these excitations will be moved away from the horizon region, leaving the later steps to again create pairs in the state (\ref{three}).  

This fact is important. People often think that the pairs created at earlier steps somehow form a `cloud' of particles around the horizon, and these particles influence all later pairs that are created.  This would indeed be the case in a normal body, but is not so for the black hole with metric (\ref{one}). We can add an initial gas of quanta to the hole, but after these quanta have stretched away, the new pair is always created by stretching of a vacuum region.

\subsection{The `large boost' argument}

Having failed in finding a suitable change to the state on the initial slice, we can try to look for changes in the {\it evolution} of this state to the next slice. The difficulty we face is of course is that our evolution region satisfies all the niceness conditions N1-N5 noted in step A of the Hawking paradox. So how do we get away from the conclusion that the evolution satisfies  (\ref{qthree}),(\ref{cond1})?

A significant number of  attempts at resolving the paradox were  based  on the failure of the Schwarzschild coordinates at the horizon. An observer at rest in the Schwarzschild frame near the horizon has a high relative velocity compared to an observer who falls in from infinity along a geodesic; this velocity approaches the speed of light as we take the observer closer to the horizon. This large `boost' between the two frames is then argued to bring in new physics, not incorporated in Hawking's original computation.\footnote{This large boost has been noted in arguments by 't Hooft and Susskind, among others.}

The difficulty with this approach is that the Hawking argument does not actually use the Schwarzschild frame at all. The Schwarzschild frame  fails at the horizon, so the argument uses instead a `good slicing' that allows us to study modes that straddle the horizon; after all these are the modes whose evolution is crucial to the entanglement process. 

We should therefore first convert the `large boost' mentioned above into an effect that can be seen on the `good slices'\footnote{This version of the `large boost' was described to me by Don Page.} and then ask what effects, if any,  this large boost  produces. 

Thus consider one of the spacelike surfaces in our foliation (fig.\ref{ftwo})). Let $n^a$ be the unit normal to this spacelike slice. In the part $r>3M$, this normal is just along the $t$ direction, while in the part $r=const$ the normal  will be in the negative $r$ direction. 

Start with this normal at one point along the spacelike slice, and parallel transport it a short distance $\delta s$  along the  slice ($\theta, \phi$ are kept fixed). The parallel transported normal differs from the normal at the new location by an amount $\delta n^a$. Compute 
$b=\Big({\delta n^a \over \delta s}{\delta n_a \over \delta s}\Big )^\h$. Finally, define the boost between two different points on the slice as
\be
B=\int_{s_1}^{s_2} b(s) ds 
\ee
 (We have used the proper length $s$, measured from an arbitrary initial point, as a parameter along the slice.)

It can be easily seen that there is no large boost along the outer part of the slice or on the `connector segment' ${\cal C}$. But there {\it is} a large boost along the $r=const$ part of the slice, because this segment is so {\it long}. 

Let us compute $B$ explicitly to see this. The unit normal along this part of the slice has nonvanishing component  $n^r=-[-(1-{2M\over r})]^{\h}=-1$ where in the second step we have put $r=M$ as the value of $r$ at which we have this part of the slice. The coordinate along the slice is $t$, with $\delta s= [-(1-{2M\over r})]^\h \delta t = \delta t$. Parallel transport gives (again setting $r=M$)
\be
\delta n^t=-\Gamma^t_{rt}n^r\delta t=-{1\over M}\delta t=-{1\over M}\delta s
\ee
Thus $b={1\over M}$, which is not large. The integral over $s$ is very large though. Each mode $c_i$ occupies a length $\delta s\sim M$ on the $r=const$ part of the slice, and the number of such modes in the evaporation process is $\sim \Big ( {M\over m_{pl}}\Big )^2$. Thus 
\be
B\sim b\Delta s \sim  {1\over M}M( {M\over m_{pl}}\Big )^2 \sim ( {M\over m_{pl}}\Big )^2\gg 1
\ee
Thus we have a large boost between the the lower and upper ends of the $r=constant$ part of the spacelike slice. The lower part is where the initial infalling matter resides (see fig.\ref{ftwo}). The upper part lies in the region where the new pairs are being produced. The crucial question now is the following: does this large boost lead to a new and  unexpected long-distance interaction between these two well-separated parts of the spacelike slice? More precisely, will  the state of a newly created pair be significantly altered  from the form (\ref{three}) due to these new effects?

And here is where the arguments invoking the large boost run into trouble: no one has shown that there is an effect in the theory of gravity that {\it will} give the desired kind of effect. The easiest way to appreciate the situation is to look at another, more familiar, example where such a large boost arises. Consider a flat Roberson-Walker cosmology
\be
ds^2=-dt^2+a^2(t)[dx^2+dy^2+dz^2]
\ee
Take the unit normal at $x=0$, which is give by $n^t=1$. Parallel transport it to $(\delta x, 0 , 0)$, and compare the transported vector to the normal at the new position. The change is 
\be
\delta n^x=-\Gamma^x_{tx} n^t \delta x=-{\dot a\over a}\delta x
\ee
Using $ds=a dx$, we find $b={\dot a \over a}$. Defining the boost $B$ just as in the black hole case, we get for two points separated in the $x$ direction
\be
B=\int_{x_1}^{x_2} b(s) ds=(x_2-x_1){\dot a }
\ee
We see that we can make this boost as large as we like by taking the points $x_1, x_2$ sufficiently far apart.

Thus we see that the large boost is present both on the Cosmological slice and on the good slice of the black hole. But in Cosmology, we do not assume that such a boost will create order unity nonlocal effects. For example, we do not expect that the evolution of a spin placed at  $x_1$ will depend significantly on  the state of a spin placed at $x_2$. In fact physics remains local in Cosmology, even though we have arbitrarily large boosts along the slice. 

Thus we see that one has to be  careful in making arguments for altering Hawking's entangled state; if we conjecture a new effect then we must first define it precisely and then check its validity in other situations in gravity.

\section{Entangling with gravity}

Hawking's argument uses no details of quantum gravity. This is justified by the nature of the good slicing where everything looks to be well described by `quantum fields on curved space', and any residual quantum gravity effects are encoded in the small corrections of step D in the argument. But there have been many attempts to argue that quantum gravity plays  a more direct role in the evaporation process.

One notion that has been sometimes mentioned is the idea that the wavefunction of matter gets heavily entangled with the gravitational degrees of freedom of the hole, and thus any semiclassical picture using `quantum fields on curved space' would miss an important aspect of the evaporation process. This idea of `entangling with gravity' sounds profound at first sight. It is also true that any wavefunction in general relativity can have such an entanglement. But our interest is in finding out if such entanglement has any consequences for the Hawking argument A-D.   In this section we will make some general remarks about `entanglement with gravity', pointing out that such entanglement is no different from any other entanglement, nor is it something that invalidates semiclassical reasoning on the good slices.

\subsection{Some suggestions about entangling with gravity}
\label{secww}

Let us first note some ideas suggesting that entanglement with the metric might be an issue relevant to the Hawking paradox:

\b

(a) For an outside observer, each emission is {\it probabilistic}; there is, say, a probability $\h$ that there is no emission and  a probability $\h$ that there is emission. Thus after an emission the black hole is left in a superposition of mass states, $M_1, M_2$. This spread in mass will grow with succeeding emissions. But the gravitational field is tied to the mass $M$ of the hole by the Gauss constraint, and so we will have a superposition of many metrics $g_i$ in the description of the system. Thus it seems inherently incorrect to describe the black hole by quantum fields on a {\it given} curved space; if we use some average metric to describe the emission then we are missing the fact that the state of the emitted quanta is somehow entangled with the state of the gravitational field.

\b

(b) Let us assume that the energy spectrum in our theory is discrete, with no degeneracies between energy levels. Then if we measure the mass of any system accurately, we can predict its internal state. In particular the mass of a black hole is known from the falloff of the gravitational field at infinity, so we should know its internal state exactly; in this sense information can never be lost in a quantum theory of gravity. 

\b

We will first make some general observations about the nature of entanglement with gravity, and then return to addressing the above arguments.

\subsection{Entangled spins}\label{seca}

Consider a spin $\h$ particle, that could be in states $|\uparrow\rangle_z$ or $|\downarrow\rangle_z$. The state with spin in the positive $x$ direction is 
\be
|\uparrow\rangle_x=\sqi\Big (|\uparrow\rangle_z+|\downarrow\rangle_z\Big )
\ee
and measurement of the $x$ spin in this state will give the value $\sigma_x=\h$ with probability $1$, the value $\sigma_x=-\h$ with probability $0$. 

Now suppose this spin is entangled with another system, which has two orthonormal states $|a\rangle, |b\rangle$. Consider the state
\be
|\psi_1\rangle=\sqi\Big (|\uparrow\rangle_z|  a\rangle+|\downarrow\rangle_z|b\rangle\Big )
\label{qel}
\ee
This time we do {\it not} make a state with spin in the $x$ direction; in fact if we measure $x$ spin we will get $\sigma_x=\h$ with probability $\h$ and $\sigma_x=-\h$ with probability $\h$. Thus entangling the spin with another system {\it changed its properties in an explicit and  measurable way}. 

Now let us consider how we might entangle with gravity. Let there be a small magnetic field in the laboratory, so that $|\uparrow\rangle_z$ has slightly more energy than $|\downarrow\rangle_z$. Then these two states will produce slightly different geometries $g, g'$.  Thus we get states $|\uparrow\rangle_z|g\rangle$ and $|\downarrow\rangle_z|g'\rangle$. Now consider
\be
|\psi_2\rangle = \sqi \Big ( |\uparrow\rangle_z|g\rangle+ |\downarrow\rangle_z|g'\rangle\Big )
\label{qtw}
\ee
Let us ask: is this a state  with spin in the $x$ direction? At first sight this state looks similar to (\ref{qel}) which was {\it not} a state with spin in the $x$ direction. But in fact (\ref{qtw}) {\it is} a state with spin in the $x$ direction. The difference from (\ref{qel}) is that whereas in that case there were {\it four} states of the overall system, now we have only two; the Gauss constraint of general relativity links $|\uparrow\rangle_z $ to $|g\rangle$ and $|\downarrow\rangle_z$ to $|g'\rangle$. The other two combinations $|\uparrow\rangle_z|g'\rangle$ and $|\downarrow\rangle_z|g\rangle$ are not allowed. In fact the operator $\hat \sigma_x$ acts as
\be
\hat \sigma_x : ~~~|\uparrow\rangle_z|g\rangle\r |\downarrow\rangle_z|g'\rangle, ~~~|\downarrow\rangle_z|g'\rangle\r |\uparrow\rangle_z|g\rangle
\ee
so that the metric created by each spin state gets `carried along' in the wavefunction at all stages. This is of course as it should be; each state in the lab has a slightly different energy and so a slightly different metric, but we do not need to write the metric along with each state because it just `gets carried along'. Thus we usually just write $(|\uparrow\rangle_z,|\downarrow\rangle_z$ in place of $|\uparrow\rangle_z|g\rangle, |\downarrow\rangle_z|g'\rangle$. To summarize, usual entanglement like (\ref{qel}) changes the way we superpose spins, but the Gauss constraint of gravity imposing (\ref{qtw})  does not have this effect.

\subsection{The wavefunction with gravity}

The wavefunction in gravity has the form $\Psi[{}^{3}g(\vec x), \phi(\vec x)]$, where we have assumed that the spatial manifold is parametrized by 3 coordinates $\vec x$, ${}^3g$ is the metric on this manifold, and the matter is taken to be a  scalar field $\phi$. The diffeomorphism invariance imposes constraints ${\cal H}_i(x)=0, i=1,2,3$ and ${\cal H}_\perp=0$ from the space and time diffeomorphisms respectively. In particular, these constraints contain the `Gauss law' for gravity -- the relation between the  the falloff of the metric at infinity and the stress-tensor in the interior.  

A general wavefunction will of course be entangled between the matter and gravity parts
\be
\Psi[{}^3 g, \phi]=\psi_1[{}^3g]\chi_1[\phi]+\psi_2[{}^3g]\chi_2[\phi]+\dots
\label{entg}
\ee
and Hamiltonian evolution will create such entangled states even if we start with a state that is unentangled between the matter and gravity parts. It is convenient to split the gravitational field into its propagating degrees of freedom  (gravitons), and the coulombic part   (which is fixed by the matter stress-tensor). The propagating modes are no different, conceptually, from  a set of scalar fields, so let us label them $\phi_k$. We can certainly entangle the matter field $\phi$ with these degrees of freedom $\phi_k$, but conceptually this is just like having several matter fields, and letting the wavefunction be entangled between them. For the coulombic mode, we had considered a simple example in (\ref{qtw}) above, and noted that the entanglement is automatic but trivial -- it just gets carried along with the matter field configuration. Thus we do not find any unexpected effects from the entanglement of matter and gravitational degrees of freedom. 

We can still ask what happens if the wavefunctions $\psi_i[{}^3g]$ in (\ref{entg}) are peaked at very different values of ${}^3g$ for different $i$. The answer is -- nothing. Each component $i$ evolves by itself, and one superposes the answer, with no serious consequences for the issue of interest to us: the entanglement between radiated quanta and the black hole. Let us examine this aspect in more detail.

\subsection{Metric `spread' in black hole evaporation}\label{secs}

Take a black hole of mass $M$.\footnote{The model discussed below was suggested by Don Page.} It emits  quanta with  $E\sim p\sim {1\over GM}$.\footnote{We restore the Newton constant $G$ in this section since it makes the estimates more explicit.} The emission of such a quantum gives a recoil to the hole with momentum $\Delta p\sim {1\over GM}$. The overall emission process emits  $N\sim \Big ( {M\over m_{pl}}\Big )^2$ quanta, and these are emitted in random directions. Thus after a significant fraction of the hole  has evaporated, the hole has a momentum of order
\be
P\sim \sqrt{N} {1\over GM}\sim  \Big ( {M\over m_{pl}}\Big ){1\over GM}\sim{ m_{pl}}
\ee
where we have used $G\sim l_{pl}^2\sim m_{pl}^{-2}$. This corresponds to a velocity
\be
v\sim {P\over M}\sim {m_{pl}\over M}
\label{qthir}
\ee
The evaporation time is $t_{evap}\sim GM  \Big ( {M\over m_{pl}}\Big )^2$. A velocity of order (\ref{qthir}) over this time would give a displacement
\be
d\sim v\, t_{evap}\sim {m_{pl}\over M}GM  \Big ( {M\over m_{pl}}\Big )^2\sim GM {M\over m_{pl}}\gg GM
\ee
Thus in the course of its evaporation the black hole moves away from its initial position by distances much larger than its Schwarzschild radius, and in general the wavefunction of the hole will involve a superposition of different positions. We can therefore wonder: if somebody used a  definite semiclassical metric (the Schwarschild metric with center at $r=0$) to study the evolution of radiation, would he come to completely incorrect conclusions?

Put this way,  this looks like an argument against Hawking's computation. But as usual, we need to follow the argument to its end to learn if we have resolved the paradox. Thus we look at a simpler situation where we also get a superposition over different center of mass positions in the wavefunction. Consider  $1$ Kg chunk of Uranium 238, containing $\sim 10^{24}$ atoms. The atoms decay slowly with a half life of  $10^{17}$ seconds.  Each decay  yields $\sim 4$ MeV of energy in alpha decay, which imparts a momentum recoil
\be
\Delta p\sim 10^{-14} gm\,  cm/sec
\ee
Emitting $\sim N$ particles in random directions gives a momentum
\be
P\sim \sqrt{N}\Delta p \sim 10^{-2}gm\, cm/sec
\ee
which implies a velocity  $v\sim  10^{-5} cm/sec$. In the decay time of $10^{17}$ sec, the chunk moves $10^{12}$ cm. This is much more than the size of the chunk, which is  $\sim 1$ cm.  

Thus we seem to have the same issue as with the black hole. The details of the evaporation are of course different: when all the atoms in the Uranium have decayed, the chunk comes to a ground state that is directly tensored with the emitted quanta, while the black hole got progressively more entangled with its radiation. But the fact that the decaying object moves  by a distance more than its own size is true in both cases. 

Now we should ask: what is the consequence of the position uncertainty for the decay of the Uranium chunk? Do we have to use any new physics which we did not use for the decay of a chunk that was fixed in position? The answer is no; each emission occurs from a location that could be different from the location at other times, but the decay process, and its consequences for entanglement, remain the same.

A similar reasoning shows that the basic relation for Hawking emission (\ref{result}) remains unchanged; the entanglement rises with each emission, even though the center of mass of the hole may slowly drift from one position at the start of evaporation to another near the end. 
Returning to our original issue, we note that the presence of many different metrics $g_i$ in a sum like (\ref{entg}) does not by itself create a novel effect that would invalidate Hawking's argument. 

\subsection{Quantum gravity and information}\label{secm}

In \cite{marolf} it was noted that because the Hamiltonian of gravity is a boundary term on shell, there may a sense in which information about the interior is coded into the gravitational field at infinity. It was further suggested that this fact may have a role to play in the information paradox: if information is always at infinity in a certain sense, then there cannot be information loss when black holes form and evaporate in the interior. Let us examine this statement in light of the Hawking argument steps A-D.  

The actual computations in \cite{marolf} relate to the algebra of boundary observables and are correct. But our interest  is in seeing what they might say about the information question.  And here we come across the following difficulty. Since the arguments are abstract, they do not tell us exactly {\it what} happened in the evolution of the black hole; they just argue that the evolution is unitary.  If we are given unitarity, then the Hawking argument forces us to a final state where the emitted  quanta are entangled with a remnant. We can take spatial infinity to be at a surface that surrounds  the entire system  -- the radiation and the remnant.  Then the argument of \cite{marolf} is consistent with the following outcome: the initial collapsing matter shell evolved to a remnant entangled with radiation. 

But in that case we have learnt nothing new about the information paradox -- remnant plus entangled radiation was always allowed as an outcome consistent with unitarity. If we wanted the other outcome -- radiation not entangled with anything -- then we had to find some effect in the quantum gravity theory that would break one of the steps A-D in the Hawking argument. But the abstract argument of \cite{marolf} does not attempt to do that; in fact it does not investigate any specific mechanism of evolution. 

To summarize, while the discussion of \cite{marolf} is significantly more sophisticated than the suggestion (b) of section \ref{secww}, at the end the issue is the same: to get the radiation to be in a pure state we need to know which of the steps A-D broke down, and abstract arguments about gravity and its Gauss law constraint do not tell us that. 

\subsection{Summary}

Some people have a general belief that `entangling with gravity'  is a relevant issue for the information problem. Since there is no concrete computation based on this belief, we have contented ourselves with exploring some ways in which entanglement with gravity can occur, and noted that in each case such entanglement does not affect the Hawking argument A-D. We  noted that the propagating modes of gravity are like any other quantum field, and so add nothing new to the issue. The coulombic mode is determined by the Gauss constraint, but as we saw by simple examples in section \ref{seca}, this effect does not really generate an entanglement in the usual sense. The idea (a) in section \ref{secww} noted that the wavefunction of the evaporating hole can be spread over widely different metrics. But as we saw by an explicit example in section \ref{secs}, such a spread has no consequences for the growing entanglement in the Hawking computation. The argument (b) of section \ref{secww} is a little naive: it does not analyze  the implications of the non-degenerate spectrum for the entanglement issue. Further, the process of Hawking evaporation  generates states that tend to be {\it degenerate} in energy. This is because the members of a created pair have opposite quantum numbers, so the two possibilities -- having the pair or not having it -- are degenerate.  The argument of \cite{marolf} had suggested that information in gravity was, in a sense, always at the boundary; this argument appears to be a more sophisticated version of the idea (b) in section \ref{secww}.  But we noted that at this level of abstraction the argument is equally consistent with remnants, and does not offer any physical mechanism that would counter the steps A-D in the Hawking argument.

\section{Subleading saddle points in the Eucliean path integral}

A few  years ago Hawking announced that he had changed his mind about the information puzzle, and now believes that  black hole evaporation is a unitary process. But people have been confused about the reasons he advanced for his change of view. His paper \cite{hawkingreverse} on the subject was not very detailed. The relativists who believed in information loss continued, for the most part, to do so.  Kip Thorne, Hawking's co-signer in the bet that black holes lost information, failed to be convinced by Hawking's new arguments and did not surrender  the bet to Preskill.

The paper \cite{hawkingreverse} does not lay out its arguments in full detail, so it is not possible to fully  accept or refute them. But we can discuss a general line of reasoning that  has been sometimes used, and that seems to be in the same spirit as \cite{hawkingreverse}. A rough version of this reasoning goes as follows. We can rotate the entire black hole evolution to its Euclidean section. There is a leading saddle point -- the Gibbons-Hawking `cigar' geometry. This much is not new. But in a full theory of quantum gravity there will be {\it other} saddle points as well. These saddle points will have a larger action and therefore be exponentially small in their effects. But these small effects might be enough to restore unitarity of the evolution, and thus avoid information loss. 

It is easy to see why such an argument would leave people puzzled. The original information paradox was formulated in Lorentzian signature. By contrast, the Euclidean section does not even have a horizon. To resolve the paradox we needed to be shown what goes wrong with the argument A-D. We can certainly take guidance from a Euclidean argument about the physics, but at the end we still need to come back to the Lorentzian `good slices' evolution and show what step in the original argument broke down.  This the Euclidean discussion did not do; hence it is not surprising the many relativists continued their belief in information loss.

Let us ask {\it why} the Euclidean argument was advanced as a solution to the information problem. If we  have a well defined path integral in Euclidean space, then we can imagine that the theory in the Lorentizian section would be forced to be unitary. Thus we would leave  no place for information loss as one of the options. The existence of subleading saddle points seems to be a subtle source of quantum gravity corrections, and one might hope that these corrections somehow resolve the problem in the Lorentzian section.

But the argument  concedes that the new effect being noted -- the contribution of  subleading saddle points  -- would be exponentially small. From the inequality (\ref{result}) we see that small corrections cannot remove the entanglement between the radiated quanta and the hole. Thus we would have to conclude that we get remnants, something that was anyway allowed by the original Hawking argument. 

The {\it hope} underlying the Euclidean argument was, however, that the small corrections would convert the Hawking radiation into an unentangled state. What led to the belief that small corrections could accomplish this task? One source of such a belief is an incorrect interpretation of a paper by Maldacena \cite{eternal}. The computations done in the paper \cite{eternal} are, by themselves,  entirely correct. These computations illustrate the fact that even in a unitary system like a CFT,  correlations in a complicated state can  become exponentially small for large time separation between the operators in the correlator. This error lies in using this fact to argue the {\it converse}, namely that exponentially small corrections remove the troublesome entanglements from the leading order Hawking state (\ref{six}). Thus, put in the context of the way we have phrased the information paradox,  the incorrect argument would go as follows: ``The leading order Hawking computation says that the radiation is in a mixed state with the remnant, but after we take into account the exponentially small corrections arising from the subleading saddle points, the radiation state could become an unentangled one. Then the hole could evaporate away and leave the radiation in a pure state.''

As we already noted, this argument is false. In fact for  the entangled state eq. (\ref{six}), we know  that small corrections do {\it not} make it a disentangled state. But to prove this, we need the inequality (\ref{result}) which is true but  not intuitively obvious since its derivation invokes the strong subadditivity of quantum entropy, a nontrivial result.

We close this discussion with a general comment. People sometimes try to phrase the Hawking paradox in terms of the behavior of correlation functions between operators at different positions. But the  paradox concerns the entanglement structure of a {\it wavefunction} on a complete Cauchy slice. While it is true that the complete information in a wavefunction can be accessed by looking at all possible correlation functions, it is also true that we cannot hope to analyze  the entanglement in the wavefunction by studying only a few simple correlators. Thus arguments based on the behavior of correlators cannot easily get to the heart of the information paradox.  The paradox is best seen in terms of the evolution of a given wavefunction on a set of complete Cauchy surfaces in a gently curved spacetime; we are then challenged to  find a physical effect that we might have have missed in this simple looking situation.

\section{What does AdS/CFT duality say about the information paradox?}

String theorists  sometimes argue in the following way: ``Many computations tell us that gravity in AdS space has a dual description as a Yang-Mills CFT \cite{maldacena}. Since the latter is manifestly unitary, the gravity theory should also be unitary, and so for black holes information should be encoded in Hawking radiation. Thus we have solved the information problem''.

This argument is, however, completely incorrect. The known agreements 
between AdS gravity and the CFT involves comparison of  scaling dimensions, n-point correlators, etc. But the Hawking argument  does not say that any loss of unitarity occurs in normal n-particle scattering. It is only when a black hole is formed that a disagreement with unitarity shows up; further, the Hawking argument purports to {\it demonstrate}   this disagreement. So the question we must ask is: what would somebody using AdS/CFT duality say about the Hawking argument? 

And here we come to the essence of the problem.  In general, a person using AdS/CFT duality  writes down a metric for black holes in AdS
\be
ds^2=(r^2+1-{M\over r^2})+{dr^2\over r^2+1-{M\over r^2}}+r^2d\Omega_3^2
\label{bhads}
\ee
which is the $AdS_5$ analogue of the Schwarzschild  metric (\ref{one}). The steps A-D on section \ref{hawking} can now be repeated. The Hawking argument forces him  to choose between information loss and remnants. What does he say now?

No one has argued that a CFT has states analogous to remnants. If there is information loss, then we lose unitarity, and the AdS/CFT map breaks down. Clearly, something is not working, so let us dig deeper into the issue.

Most string theorists believed that the metric (\ref{bhads}) was correct.  The good slicing  holds for this metric (step B of the Hawking argument).   At this point there is a divergence of opinions. Some believed that small corrections due to quantum gravity would modify the Hawking argument enough to restore unitarity, but with the inequality proved in \cite{cern} (step D of the argument) we know now that small corrections cannot help. Others believed that there would be a `complementarity' \cite{suss1} which would allow information to be both outside the hole and inside, but how do we show any such thing? The state on the good slice is no different from the state on any normal slice in gravity;   the horizon location does not show up as a special place at all. How can we duplicate the information on one part of this slice and place it on the other part, without a similar rule being applied to every partition of every slice in gravity? 

The simplest way out would be if the black hole was {\it not} given by the metric (\ref{bhads}); then the Hawking argument would not proceed (we would break step B). But this is equivalent to finding `hair' -- a modification of the black hole solution. And now we are back to square one: if we knew how to construct hair for black holes, we would have already removed the paradox, with no involvement of AdS/CFT. 

\subsection{Defining gravity as the dual of a CFT}

Some string theorists try to use a slightly different argument. It is of course agreed that the CFT is unitary. Let us {\it define} the gravity theory as the dual of this CFT. Then surely we cannot have any paradox? 

\begin{figure}[htbp]
\begin{center}
\includegraphics[scale=.30]{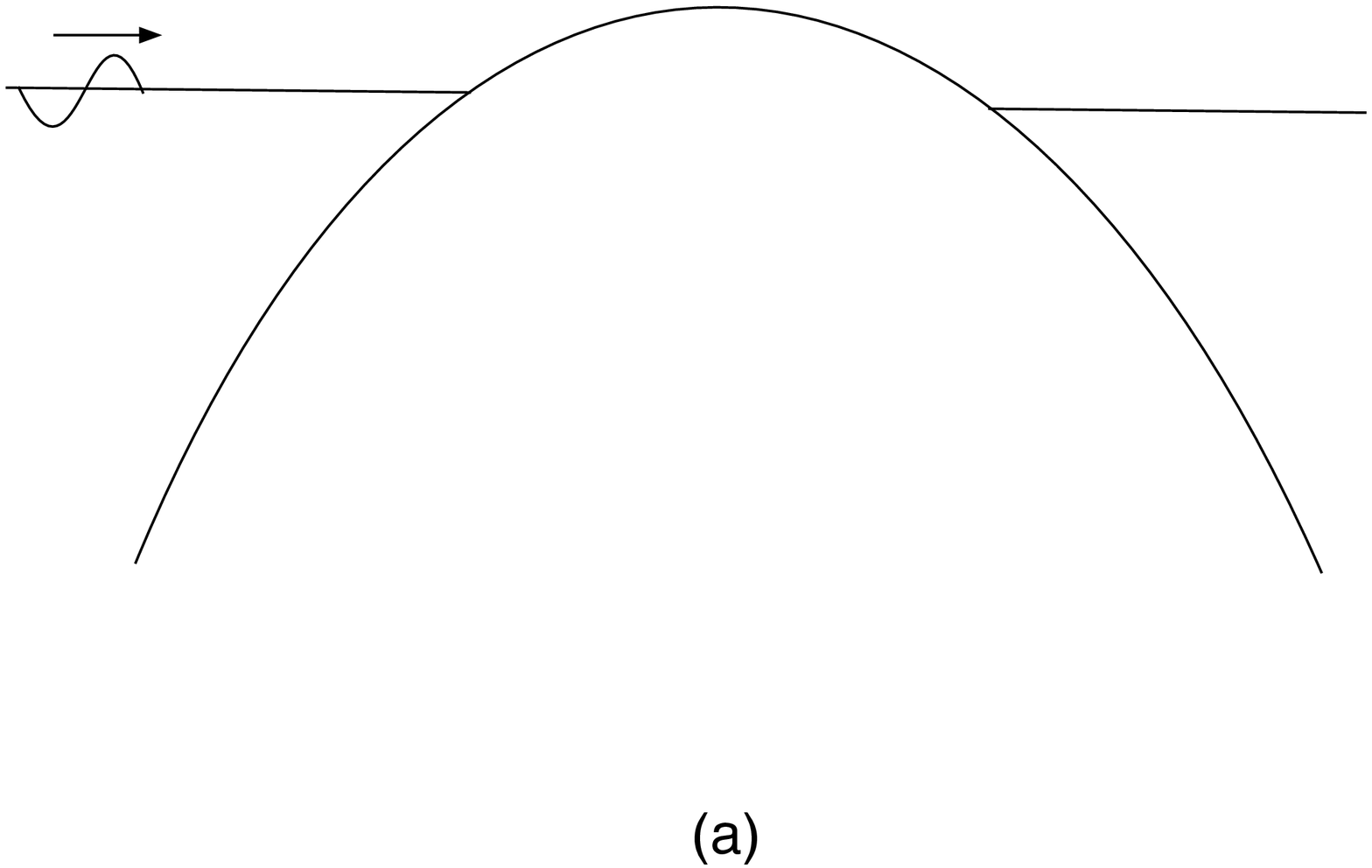}\hskip 1.0 truein\includegraphics[scale=.30]{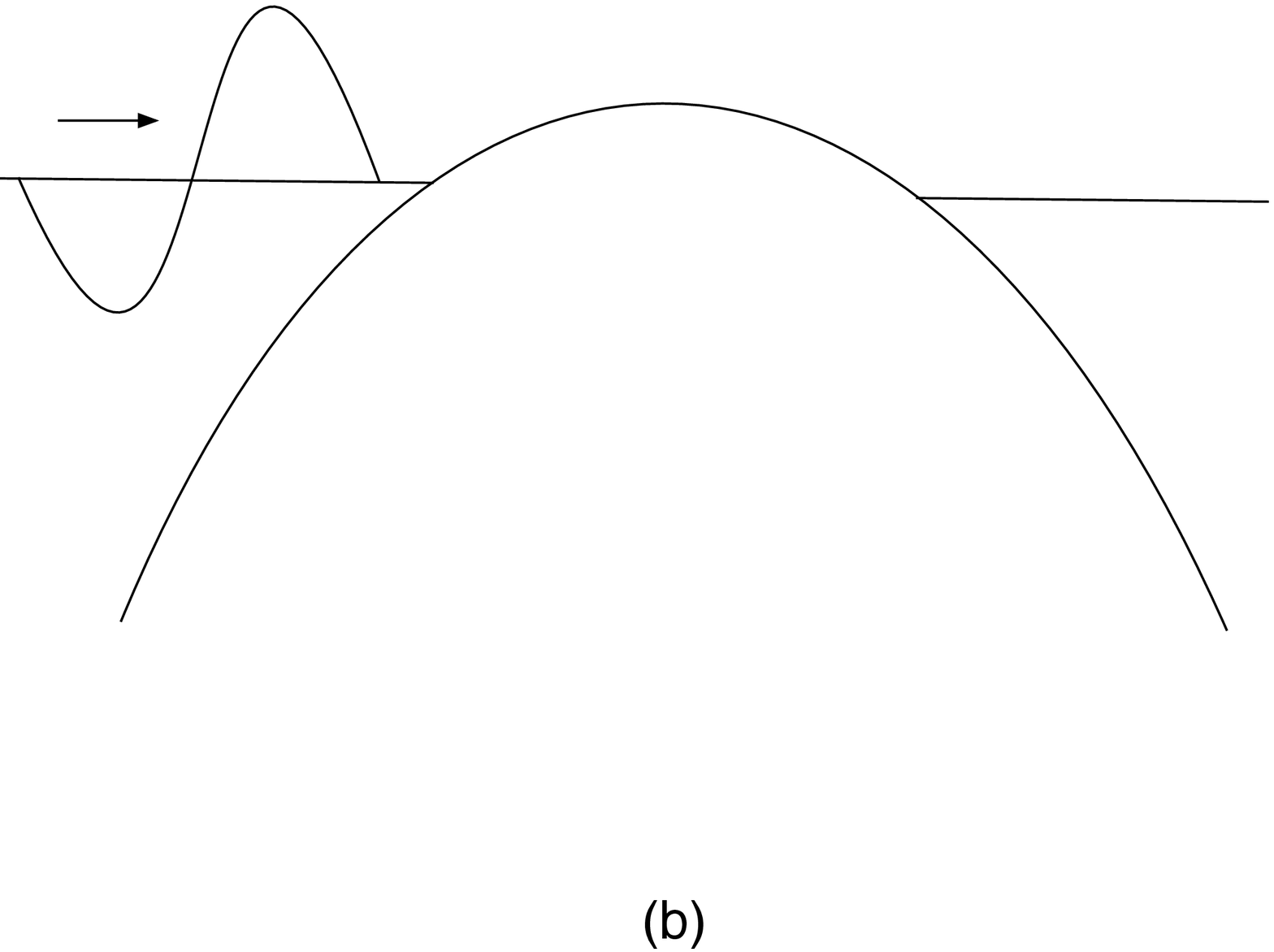}
\caption{{The horizontal direction labels the value of $\lambda$, the eigenvalue. The horizontal line gives the level of the fermi sea of eigenvalues, obtained when the matrix model potential is the indicated `upside-down harmonic oscillator' potential. (a) A small pulse representing a quantum scatters off the wall at $r=0$ (b) A pulse large enough to make a black hole in dilaton gravity spills over the wall instead of returning to the same asymptotic infinity. }}
\label{ffive}
\end{center}
\end{figure}

In fact, this argument is equally empty, but it is very illustrative to examine where it goes wrong. We have seen something like AdS/CFT before, in the solution of the $c=1$ matrix model. (This matrix model  in some sense offers a  simple version of the AdS/CFT duality map.) Strings can be exactly quantized in 1+1 dimensions, and the result mapped onto a large N matrix model in  0+1 dimensions \cite{dj}. The relation between the  two descriptions is obtained by looking at the eigenvalue distribution of the matrix. The eigenvalues form a 1-dimensional `fermi sea', characterized by the density of eigenvalues $\rho(\lambda)$. In fig.\ref{ffive} we plot the potential in which the eigenvalues live. The depth of the sea at any point gives the density of eigenvalues $\rho(\lambda)$.  We can distort the eigenvalue distribution (the small ripple shown in fig.\ref{ffive}(a)), whereupon it evolves in time to give a 1+1 dimensional function on the space $(\lambda, t)$. The dynamics of such ripples gives string theory in 1+1 dimensions, and its low energy limit is just 1+1 dimensional dilaton gravity.  Ripples come in from spatial infinity (the extreme left of the fig.\ref{ffive}(a)), bounce off the potential barrier and return to infinity. In the gravity theory, this is like a quantum coming in from $r=\infty$, hitting $r=0$ in flat space, and going back to $r=\infty$.

Correlations functions in the matrix model are found to reproduce the expected correlation functions for n-particle scattering in the gravity theory. In particular,  the matrix correlators reproduce the all-important  `time delay' caused by the couloumbic gravitational attraction of one  particle on another \cite{pn}.   Thus, so far, the situation is just like the computation of n-point correlators in AdS/CFT, where we also find that the CFT reproduces the expected behavior of low energy gravitational scattering.   But what about black holes?

As mentioned above, the low energy theory of the matrix model is just 1+1 dimensional dilaton gravity. In dilaton gravity we have black holes,  with a Penrose diagram similar to the Penrose diagram of the usual Schwarzschild hole.  Hawking radiation in dilaton gravity has been studied in great detail, and the 1+1 dimensional case allows a very explicit computation of the entangled Hawking state \cite{giddingsnelson}. One finds that the evolution ends in  information loss/remnants \cite{cghs}. 

But if  we have started with the matrix model, and obtained  dilaton gravity as a low energy limit, then in the full theory we do not expect information loss.   But there seems to be no place for remnants in the matrix model either. So what is going on?

The answer is simple: the 1+1 dimensional theory gravity dual to the matrix model does not have black holes \cite{pn,kms}. It is true that n-point functions in this gravity theory behave just as one would expect n-point functions to behave in gravity. These functions are computed by the scattering of perturbative ripples on the fermi sea. But to make a black hole we must make a large wave, not a perturbative ripple, and see how it evolves. As we can see from the picture of the fermi sea in fig.\ref{ffive}(b), a large wave approaching the bump at $\lambda=0$ simply spills over to the other side. If the energy does not return to the same asymptotic infinity from which it started, then we should say that we have information loss, even though the overall theory was unitary. We can make a $Z_2$ identification of the two sides of the fermi sea -- this is automatic in some supersymmetric versions of the string theory -- and then the energy returns to the same asymptotic infinity from which it was sent. But most of it  returns in a very short time -- what we would call the `crossing time' in gravitational collapse. In a black hole, on the other hand,  the return of energy should be through a slow Hawking radiation process, over a  time which is  some power of ${M/m_{pl}}$ times the crossing time. Thus with the orbifolding we get unitarity but no black hole.

Instead of orbifolding the fermi sea we can try to change the matrix model potential in other ways, for example by putting an infinite spike at $\lambda=0$ so that there is no spillover. But such corrections make unwelcome changes to the n-point gravity correlators; in general they they can no longer be derived from a local gravitational theory. 

Thus the matrix model example leads us in an interesting dance all around the Hawking argument. In the simplest version with two sides to the fermi sea, we  have a version of information loss -- energy flowing off to a second  asymptotic infinity. If we block this avenue by $Z_2$ orbifolding, we do not get anything that behaves like a black hole; thus we violate step B in the Hawking argument. If we try other changes to the potential, we lose locality, violating step A. What we do not get is a hole with a traditional Penrose diagram, slowly emitting radiation in a unitary way.

Let us return to AdS/CFT and ask what  we can say in that case. The n-point correlation functions are computed a $1/N$ expansion. But to make a black hole we need $\sim N$ quanta to interact, so we have to look out for unexpected effects. In the matrix model case, it is the large number of interacting quanta which led to deviation from dilaton gravity.  Any perturbative amplitude in the matrix model has higher order corrections that make it deviate slightly from dilaton gravity. But when the wave is big enough to make a black hole then we get a {\it large} correction because we have interactions  between the large number of quanta in the wave. The cumulative effect of these corrections changes the evolution to one different from the one expected in dilaton gravity, and the black hole does not form at all. Returning to the AdS/CFT case, if we force the gravity theory to be the dual of a given CFT, then we cannot assume that black holes will form in the theory. 

To summarize, simply invoking the idea of AdS/CFT duality does not give us any useful statement about the information paradox. The paradox asks us to demonstrate a flaw in the Hawking reasoning. But just noting that the CFT is unitary does not tell us how to resolve the paradox.  We would still  not know how to choose among the various options: 

\b

(a) the duality map itself breaks down upon black hole formation 

(b) there are long-lived remnants in both the CFT and the gravity descriptions

 (c) black holes do not form in the gravity theory
 
  (d) long lived `black holes' form upon gravitational collapse but are not given by the solution  (\ref{bhads}). 
 
 \b
  
  Option (d) would allow information recovery in Hawking radiation, but to get this option we have to find `hair': states of gravity that do not settle down to the metric (\ref{bhads}). But if we knew how to construct such hair in the gravity theory, we would have no paradox in the first place since step B in the Hawking argument does not have to be true. So the paradox should be tackled by looking for alternative solutions in the gravity theory. The dual CFT description may help us in this search (and it certainly does), but resolution of the  paradox has nothing to do with the existence of any such dual description.

\section{Fuzzballs}

We have examined some attempts to resolve the information paradox and noted in each case that one or more of the steps in Hawking's argument were not adequately addressed. What then is the resolution of the paradox?

Hawking's original argument (steps A,B,C in section \ref{hawking}), supplemented with step D, leaves us with very little choice. If we have a black hole with its standard Penrose diagram, and the normal vacuum state at the horizon, then we will say that we have a `traditional black hole'. There is no `data' on the horizon since this region is locally the vacuum; we call this an `information free horizon', and propagation of modes in the vicinity of this horizon is well described by quantum fields on gently curved space. The theorem of section \ref{hawking} tells us that with a traditional black hole we {\it cannot} solve the problem: we will have information loss or remnants.

We use the term `fuzzball' to signify the opposite of the traditional black hole.  In the traditional black hole there is a good slicing with an information free horizon. In a fuzzball,  we {\it cannot} find such a good slicing with an information free horizon. At this point, the term `fuzzball' is just a definition. But it is clear that if it turns out that in  our theory of gravity  black holes are indeed fuzzballs, then there is no information paradox, since the geometrical structure of the hole used in the Hawking argument becomes invalid.

It may seem that we have been overly broad in defining `fuzzballs'; we have just said that whatever is not a traditional hole is a fuzzball. But this is exactly what the `no-hair theorem' suggests we do. Considerable work with black holes had failed to find any time independent deformation of the standard Schwarzschild geometry. Thus it is logical to make the dividing line between the traditional Schwarzschild hole and `everything else', which is what we have done with our definition of fuzzball. 

Of course we have not solved anything with this definition; we need to go to our theory of gravity and {\it show} that black holes are really described by fuzzballs and not by the traditional Schwarzshild geometry. In this section we will summarize some work in string theory where we  construct states of black holes and observe that they do not have a traditional horizon; i.e., they are fuzzballs.

\subsection{The nature of fuzzballs}

The traditional approach to black holes starts by writing a metric ansatz
\be
ds^2=-f(r) dt^2+g(r) dr^2+r^2d\Omega_2^2+dz_idz_i
\ee
where we have assumed that the metric coefficients are independent of the angular coordinates and also any compact coordinates $z_i$. Einstein's equations then lead us to the traditional hole with `information free horizon'.  We can attempt to put back dependence on the angular variables or the compact directions by looking for small perturbations to the traditional hole; the `no-hair theorem' is the statement that one finds no such finite energy deformations.  

It turned out however that there are {\it non-perturbative} deformations that {\it do} give a complete set of `hair' for the hole, i.e. one state for each of the $Exp[S_{bek}]$ states of the hole. The structure of these states involves the compact directions and the angular directions in a nontrivial way. Consider one compact circle $S^1$. Fibering this circle over the angular direction can give a KK-monopole. Since the hole has no net KK-monopole charge, we must have an anti-KK monopole somewhere else, or more generally, a set of monopoles and antimonopoles with net charge zero. In the simplest case studied -- the 2-charge extremal hole -- one gets a KK-monopole times a spatial loop; opposite sides of this loop have KK-monopoles in opposite orientations so that the overall monopole charge cancels out \cite{fuzzball1}. The space of different loop shapes can be quantized and yields the correct entropy for the hole \cite{rychkov}. 

This case is particularly simple -- the 2-charge extremal hole is called the `small black hole' in string theory -- but it is  illustrative because it can be solved very explicitly. In string theory, the states of an extremal system (mass=charge) can be counted by an indirect method: the number of extremal states do not change with coupling so they can be counted at weak coupling where there is no gravity. Suppose one then goes to stronger coupling where a black hole is expected, takes the spherically symmetric ansatz, and computes the Bekenstein-Wald entropy from this geometry. The state count at  weak coupling is found to agree with the Bekenstein-Wald entropy of this spherically symmetric solution \cite{sen,dabholkar}. But if this spherically symmetric solution were the correct story, then we would still be left with  the information puzzle, since we do not have `hair'; if we excite this system and let it radiate back to extremality, we would create the entangled Hawking pairs.   

It turns out however that  in this simple 2-charge system it is possible to study the states from first principles and realize that the spherically symmetric ansatz is {\it false}; each of the states counted at weak coupling can be constructed at strong coupling and {\it observed} to have the `KK-monopole times a loop' structure mentioned above. The only way to make a spherically symmetric state is to take a linear superposition of non-spherically-symmetric solutions (with equal amplitude for each orientation).  

Interestingly, the `KK-monopole times loop' geometry does not have a horizon; nor does it have a singularity when we include all the dimensions including the $S^1$ which was fibered to make the KK-monopoles. (If we dimensionally reduce on this $S^1$, the reduced solution will have singular metric coefficients at the centers of the KK-monoples.) Making the loop progressively more convoluted takes us to more and more generic states of the system, till the quantum measure cuts off the wiggles of the loop at some scale to yield a finite state count that reproduces the entropy. Since the generic state comes from very wiggly loops, it is very `quantum fuzzy', and so the solutions are termed fuzzballs.

\subsection{The expansion in `complexity'}

One may wonder why such fuzzball solutions were not found earlier, given how much effort was put into the search for `hair'. When looking for new solutions one often searches for a small parameter to expand in. In general relativity the only expansion parameter that appeared relevant was the amplitude of the perturbation, and this would not yield the KK-monoples since  monopoles are an inherently  nonperturbative deformation.  Nevertheless,  KK-monopoles have been known for many years to relativists, so we should still ask: what new tool did string theory bring to bear in the search for hair?

What string theory did was provide a new expansion parameter -- the {\it complexity} of a microstate. In string theory we can understand all states as bound states of various objects in the theory. We can categorize  these bound states  in a manner similar to  the states of electromagnetic radiation in  cavity. The simplest state of electromagnetic radiation has all the energy in one fourier mode -- this is a `laser beam' state with very low quantum fluctuations, well characterized by a classical solution of Maxwell's equations. Next, we can split the energy among two modes, getting a state that has slightly more fluctuations since the number of photons per mode has  decreased. Continuing this way we arrive at the generic states of the system, the typical blackbody radiation states, which have $\sim 1$ excitations for each typical occupied mode, and thus no good description in terms of Maxwell's equations. 

The situation is the same for black hole microstates. Looking at the set of all bounds states in front of us, we find one that can be characterized as having  `all quanta in the same mode'. We can explicitly construct the gravitational solution of this state, at the coupling where the black hole was supposed to exist. We find the `KK-monopole times a loop' structure, with the loop in this case being a geometric circle. There is no horizon or singularity.

This state by itself is not a generic one -- it was actually the most `ungeneric' one -- so even if we had come across it in studying the solutions of gravity we might not have thought of it in connection with the black hole. But we can now move to more generic bound states in string theory, and find that the shape of the loop becomes progressively more convoluted. When we reach the generic state the convolutions are at planck scale, and quantum effects are order unity, but the main point is that we observed the {\it path of approach} to the generic state through a family of states that {\it do not have a traditional horizon}. The states obtained in the generic limit are a `quantum mess', but someone trying to follow Hawking's argument can no longer claim that the state of the black hole is given by a `traditional horizon': if the states we  can understand explicitly have no horizon, then there is no reason why quantum corrections should make the more complicated ones {\it have} a  horizon. Instead of an `information free horizon' we get a region with quantum fuzz containing the details of the microstate. We thus knock out step B in the Hawking argument  and  resolve the information paradox.

\subsection{Three charge extremal holes and nonextremal holes}

We can make more general extremal holes in string theory by combining three kinds of charges -- the 3-charge extremal holes.   The count of bound states again  agrees with the Bekenstein entropy of what we will call the `naive geometry' -- the black hole obtained by assuming a spherically symmetric ansatz for the given mass and charge values \cite{sv}.\footnote{The papers \cite{larsen} obtained a microscopic count of states that was very similar to \cite{sv}, though the numerical factor did not precisely agree with the value expected in string theory. These papers also noted that the information problem should be resolved with a construction of a full set of `hair' for the black hole.} When we start by looking at the structure of individual microstates, starting with the simplest, we again get solutions that have no horizon or singularity \cite{fuzzball2}. Large families of such regular solutions have been constructed with  the same quantum numbers as the hole, and the entropy of such no-horizon solutions has been argued to be of the order of the Bekenstein entropy. 
Thus it seems reasonable to expect that the 2-charge extremal story will repeat itself for the 3-charge extremal case. Similar solutions are also found for the 4-charge extremal case, which gives black holes in 3+1 noncompact dimensions.

We can also look at simple {\it nonextremal} string theory bound states, and in some cases their gravity solutions have been constructed as well \cite{ross}. The solutions again have no horizon or singularity, but they do have an {\it ergoregion}. Ergoregions radiate quanta by the process of ergoregion emission, with one quantum radiating off to infinity and another falling into the ergoregion.  For the nonextremal microstate of \cite{ross}, the rate of leaking energy was computed in \cite{myers}. It turns out  that this rate of leakage agrees exactly with the expected Hawking radiation from this (non-generic) microstate \cite{cm1}. There is no information loss problem however, since the infalling member of the pair created in ergoregion emission just resides in the ergoregion, and influences the creation of later quanta, just as would be the case in a model of `burning paper'. 

\subsection{Dynamics}

The fuzzball construction aims to find all energy eigenstates in string theory for given total values of mass and charge.  In any quantum system the dynamics of any state is known once we know the eigenstates
\be
|\psi(t)\rangle=\sum_k e^{-i E_k t} |E_k\rangle
\ee
Thus in principle we  have have already learnt all we need to know about the gravitational collapse of a massive shell.  But from this formal expression  we would like to extract a qualitative answer to the following  basic question: if we have gravitational collapse, how do we get anything other than the standard Penrose diagram of the Schwarzschild metric? Where does the semiclassical evolution go wrong, given that the curvature appeared to be low everywhere except near $r=0$? 

The collapsing shell is a superposition of energy eigenstates; otherwise it would be a time independent configuration rather than a `collapsing' shell. In traditional relativity the states we superpose correspond to the shell at various radial positions -- a 1-parameter family. In that case the evolution of the shell looks straightforward; one simply gets radial motion all the way down to $r=0$. The resulting geometry gives rise to the information paradox.

Let us see how fuzzballs change the situation. The fuzzball construction forces us to recognize that there are $Exp[S_{bek}]$ {\it other} solutions in gravity that have the same overall quantum numbers. In general the wavefunction of the shell can spread over all these states. The  question now is: will such a spread be significant enough that we should think about it?

One way to estimate the spread is to take the state of the shell and ask for tunnelling amplitude to one of the fuzzball states. This amplitude is very small, since we are asking for the action to transition between two different macroscopic configurations. This smallness is the reason why we do not worry about the existence of alternative solutions in macroscopic physics -- we just focus on the classical evolution path. But in the present case there is an important difference: the {\it number} of alternative states -- $Exp[S_{bek}]$ is very {\it large}. A simple  estimate  finds that the smallness of the tunnelling amplitude is just offset by largeness of the number of states that we can tunnel to \cite{tunnel}. The time required for the wavefunction of the shell to spread over the fuzzball eigenstates was estimated in \cite{rate}, and found to be much less than the Hawking evaporation time. Thus we have an interesting  situation: the evolution of the collapsing shell is not classical at all. The measure in the path integral is as important as the classical action, due to the very large number  ($Exp[S_{bek}]$) of possible states. The wavefunction of the collapsing shell spreads out over the space of fuzzball states before a significant part of the evaporation has proceeded, and we lose the `information free horizon'. 

\section{Some common confusions about  fuzzballs}

Resolving the information paradox only needed us to find an order unity correction to the evolution (\ref{six}), so that the Hawking argument could be invalidated. The fuzzball construction does much more, giving us explicitly the structure of specific black hole microstates, and for simple nonextremal ones, even showing the Hawking radiation as an explicitly unitary process. 

There have been however, some objections to fuzzballs as the correct description of black holes. In this section I will try to collect these objections into the form of a discussion; thus I will join together the objections raised by different people into a continuous thread. As we will see below, the objections can be finally traced to an incorrect understanding of either the information paradox or the fuzzball construction. 

A primitive form of the complaint is the following:  ``Your fuzzball solutions are classical metrics. A state of a black hole must be a quantum state. So how can fuzzballs have anything to do with microstates of  black holes?''. 

It turns out that this question arises from a confusion about how classical solutions arise as a limit of a quantum field; thus it has nothing to do with black holes per se.  I have found that the best answer is to ask a  question in return: ``We write the metric of global AdS as the dual of the ground state of the CFT. How can the classical metric of AdS have anything to do with a quantum state of the CFT?"

This `counter-question' usually clarifies the issue.  A solution to the classical field equations should be thought of as giving the peak of a  Gaussian wavefunctional. This wavefunctional is typically a coherent state of the full field theory. If we like, we can take combinations of coherent states and recover energy eigenstates. 

With this simple point cleared, let us move on to the next level of the compliant, which goes as follows:  ``The simpler fuzzball states are well described by giving the metric, but the generic ones are very complicated, so we cannot ignore quantum corrections. If we do not give the full quantum wavefunctional of these states, we have not said anything useful about them. Since the black hole will be in one of these generic states, we have not said anything about the information paradox.''

This complaint also has no merit. String theory allows us to  list all states with given quantum numbers. Starting from the simplest we have constructed them and found them to be fuzzballs. First  order quantum corrections have also been computed and shown to not alter the fuzzball structure \cite{higher}. Nobody has given a reason why  quantum corrections should make the more complex states  change their horizonless structure and develop an `information free horizon'. Further,  for the simplest hole (the 2-charge extremal hole) {\it all} states are known to be fuzzballs.  All we had to do to resolve the paradox was to show a path by which one step in the argument can break down, and have we not shown that this fuzzball structure removes a reason for accepting the metric (\ref{one}), and thus invalidates step B?

This would seem clear, but for a few people, the complaint has persisted. To understand the complaint better, I posed a hypothetical question.   Suppose we go back to the early days of the paradox, and look at the computations where people looked for `hair'. They took a scalar field $\phi=Y_{lm}(\theta, \phi)e^{-i\omega t} f(r)$, and looked for solutions for $f(r)$. No finite energy solutions were found, but suppose for the sake or argument that we {\it did} get a good solution for each $\{ l, m\}$. When $l$ gets so large that the angular wiggles are order planck length, the modes will presumably be cut off by quantum gravity effects. We would have one mode per planck length squared of the horizon, so we have the right number of degrees of freedom to account for the entropy. Since the black hole can have these degrees of freedom excited, it does not have to have the vacuum state at the horizon, and the Hawking argument fails. Would we have solved the information paradox in that case?

The people who complained about fuzzballs said no, this would {\it not} solve the paradox.  They argued that since the typical state would have large $l$, quantum gravity corrections would modify the leading order solution for the hair, and it would be unclear what the nature of the generic hair state would be. Thus we would not have proved that evolution in the black hole was unitrary; we need the precise quantum gravitational nature of the generic hair state to be able to compute the evolution of the hole and check that it is unitary.

This discussion showed that the complaint against fuzzballs  had nothing to do with fuzzballs per se; it was a confusion about what the information paradox {\it is}. The proponents of this compliant believed that information would come out in the Hawking radiation anyway; either by small corrections to Hawking's computation or by large corrections to the black hole metric. They did not realize that small corrections {\it cannot} work, and they did not realize that large corrections to the black hole metric had been long sought but not found (the no-hair problem). So they just {\it assumed} that Hawking radiation would be in an unentangled state at the end of evaporation, just as it would be from a piece of burning paper. Having made this assumption, they were looking for a derivation of the exact S-matrix, and so were asking for a full quantum description of the generic fuzzball. 

But as we have seen, the Hawking paradox is something else; it is a concrete  claim for information loss, which argues that black holes do {\it not} behave like burning paper. To resolve the paradox  all we have to do is present evidence that one of the assumptions in incorrect, and this the fuzzballs do.

\subsection{Collective modes of fuzzballs}

We should still ask ourselves {\it why} the above confusions about fuzzballs arise. Pushing a little further, one finds the following question: ``The generic fuzzball is a very messy and fluctuating quantum state. The vacuum of quantum gravity is itself a messy and fluctuating state at the planck scale. So as you go to the generic fuzzball state, have you not gone back to an effectively information free horizon?''

The answer is {\it no}, a messy fuzzball state is {\it not} the vacuum. If we start with  two states that are orthogonal at leading order, and  add quantum corrections to both, then they remain two orthogonal states; they do not turn into the `same' state. 

But one can push the question further: ``Even if the generic fuzzball state is orthogonal to the vacuum, and thus not close to the vacuum itself, could it not be true that such generic states behave very much like the vacuum for all practical purposes?''

And here we come to the essence of the matter: for the information problem, there is only {\it one} process with any practical purpose:  the slow leakage of radiation from the microstate. But for this process the answer is already clear. We have seen explicitly how radiation emerges from simple nonextremal microstates -- it comes from ergoregion emission.  For the simplest microstate the ergoregion is very simple -- it is an exact circle -- and the emission spectrum is peaked at a few sharp frequencies. The radiation rate $\Gamma(\omega)$ from the ergoregion  is found to agree exactly with $\Gamma_{CFT}(\omega)$, the emission rate computed for that microstate in the CFT; the sharp peaks in the $\Gamma_{CFT}$ arise from the fact that all excitations are `in the same mode' for the simplest microstate.  As we move to a more complicated gravity solution,  the ergoregion becomes more `wiggly'. It can be seen qualitatively that  such an ergoregion will  radiate more slowly with a more smeared-out spectrum. The same behavior holds for the corresponding CFT states, where the excitations are now spread out over more modes.  The generic gravity solution will correspond to the generic CFT state. The generic CFT state has an emission spectrum that agrees with the spectrum of {\it Hawking} radiation. Thus we expect that the generic, messy gravity states radiate with the Hawking radiation profile (though without information loss).\footnote{In particular, note that fuzzballs are different from the `massive remnants' proposed in \cite{giddingsrem}. Such remnants are {\it not} expected to radiate with a spectrum related to that of black holes.}

Thus for the purpose of Hawking emission we have obtained exactly what we wanted: different microstates do {\it not} behave the same way; the details of the microstate get encoded in the slowly leaking radiation. But we can still ask: ``What about other physical processes, as for example the infall of a rocketship into the black hole? Do all generic microstates behave the same way when impacted by a heavy, fast falling object?''

This time the answer is expected to be {\it yes}. The crucial distinction therefore is between (i) low energy processes like Hawking emission, which have $E\sim T$ and (ii) high energy processes with $E\gg T$. High energy impacts excite collective modes of the fuzzball. These modes are insensitive to the precise details of the microstate, and one can hope to get a  `membrane paradigm' \cite{membranebook} type of behavior from the fuzzball surface \cite{membrane}. Some discussion of these issues can be found in \cite{bala, plummat}, though much remains to be understood. But for the purpose of the information paradox the story is clear: different microstates behave differently for the purpose of Hawking emission, and this is what resolves the  paradox. 

A final note on this topic. A very common statement is the following: ``We know that nothing happens at the horizon of a black hole, since I would not know if at this moment I was falling through the horizon of a very large hole. Thus I cannot accept that black holes are fuzzballs, since fuzzballs do not have a regular horizon.''

It is amusing that the {\it question} is misguided though the {\it answer} is still interesting. Why did one assume that nothing happens at the horizon of a large hole? Presumably because of our reliance on the metric (\ref{one}). But we have already seen that  this metric would force us  to information loss/remnants, and also that in string theory this is {\it not} the metric of black hole microstates. We must accept the metric (\ref{one}) if we assume (as Hawking did) that quantum gravity can be relevant only when curvatures reach the planck scale $l_p$. It is certainly true that $l_p$ is the only length scale we can make from $c, \hbar,G$. But a black hole is made of a large number of quanta $N$, and we have to ask if the size of string bound states remains $\sim l_p$ when $N$ is large, or whether it grows as $\sim N^\alpha  l_p$ for some $\alpha>0$. String theory has taught us that the latter is true, and the size of brane bound states is always order horizon size \cite{emission}, so that at large coupling we get fuzzballs instead of black holes. 

Thus there was no reason for us to assume a priori that the black hole would have no structure at its horizon radius: this is a matter of computation in a given theory of gravity, and in string theory we find that there is indeed nontrivial structure at the horizon scale. What is interesting though is that  the infall of $E\gg T$ objects excites collective modes of the fuzzball that might {\it approximate} `free-fall' through a horizon. This effective free-fall behavior  is suggested by putting together  arguments put forth by Israel \cite{israel}, Maldacena \cite{eternal} and Van raamsdonk \cite{raamsdonk}, and is discussed in \cite{plummat}. 

To summarize, it is crucial to separate $E\sim T$ physics from $E\gg T$ physics. The former sees structure at the horizon, and this resolves the information paradox. The latter sees a `coarse-graining' of this structure, and yield physics that might accord with intuition about classical infall into black holes.

\section{Discussion}

The black hole information paradox has played a  remarkable role in physics. Normally we expect a separation of scales, so that the quantum theory and gravity would be connected only when we reach planck distances. But the paradox forces us to confront the peculiarities of quantum mechanics -- entanglement in particular -- in the context of astrophysical sized gravitating objects.   

 The paradox is best phrased in terms of a {\it wavefunctional} on a good slice, and its evolution to the next slice. These slices look just as smooth as slices through a laboratory on earth. There are only two possibilities: (i) low energy evolution on these slices is `lab type' evolution upto small corrections (ii) low energy evolution differs from lab evolution by order unity. In  case (i) we necessarily get information loss/remnants. In coming to this conclusion the inequality (\ref{result}) plays a crucial role. Many people had wanted to keep the evolution of (i) but remove the Hawking entanglement  through the cumulative effect of small corrections; with (\ref{result}) we now know that those efforts were bound to fail. If we wish to avoid information loss/remnants we need to be in case (ii), but here we run into the no-hair `theorem': despite considerable effort people were unable to find a structure of the black hole that would {\it not} have `lab evolution' on the good slices.

Since string theory appears to give us a complete  theory of quantum gravity,  there has been a strong tendency to just wish the paradox away. But this is not easy, as we can see from the fact that many prominent relativists remain unconvinced that information loss can be avoided. Worse, string theorists do themselves a disservice by not understanding the paradox: if the usual assumptions of physics are leading to a problem, then there is sure to be a deep and important lesson to be learnt from resolving the problem.

And indeed, we do learn something very basic by looking at how fuzzballs finally resolve the paradox. The remarkable thing about black holes is their very large entropy -- $Exp[S_{bek}]$ -- but the metric (\ref{one}) shows no evidence of the large  degeneracy of microstates implied by this entropy. In string theory we find that the metric (\ref{one}) is not realized in the theory, but there  are $Exp[S_{bek}]$ {\it alternative solutions}, each without any regular horizon. The phase space of these solutions leads to a path integral measure so large that it competes with the classical term in the action, even though we are looking at the collapse of an astrophysical sized shell. The classical infall of the shell gets modified, and we end up with a linear superposition of fuzzball states rather than the solution (\ref{one}). 

The key lesson here is that gravitational solutions in a complete theory of gravity have an enormous phase space, and that this fact must be taken into account whenever enough energy accumulates in a region to make this phase space accessible. It is plausible that this lesson would be relevant in the early Universe, where we encounter high densities of matter. It was noted in \cite{cmuniv} that a `fractional brane gas' state of matter has more phase space at early times than a conventional `radiation dominated' phase. The phase space available to these fractional branes grows rapidly with the volume available to them; in particular, one gets more phase space as the horizon radius expands. 
This extra phase space can create a `pull' towards larger scale factor values, supplying a `measure term' in the effective dynamics of the scale factor that would be analogous to the measure term that alters the dynamics of a shell collapsing to make a black hole. This extra `pull' can resemble  an extra `acceleration term' in Cosmological expansion; we hope to return to this issue elsewhere. 

\section*{Acknowledgements}

These notes are a consequence of many discussions with many people; among them  Steve Giddings, Gary Horowitz, Werner Israel, Don Marolf, Emil Martinec, Shiraz Minwalla, Ashoke Sen and Edward Witten. My special thanks to Don Page, who spent many hours with me examining different ideas on the paradox, and suggesting models and examples. This  work was supported in part by DOE grant DE-FG02-91ER-40690.

\end{document}